\documentclass[fleqn,usenatbib]{rasti}

\usepackage{newtxtext,newtxmath}

\usepackage[T1]{fontenc}

\DeclareRobustCommand{\VAN}[3]{#2}
\let\VANthebibliography\thebibliography
\def\thebibliography{\DeclareRobustCommand{\VAN}[3]{##3}\VANthebibliography}

\usepackage{graphicx}	
\usepackage{amsmath}	
\usepackage{pdflscape}
\usepackage{xcolor}
\usepackage{longtable}
\usepackage[width=\textwidth]{caption}
\usepackage{placeins}
\usepackage{hyperref}

\newcommand{\Teff}{\mbox{$T_{\mathrm{eff}}$}}
\newcommand{\Ion}[2]{#1{\,\sc#2}}
\newcommand{\logX}[1]{\mbox{$\log(\mathrm{#1/He})$}}
\newcommand{\logXY}[2]{\mbox{$\log(\mathrm{#1/#2})$}}

\newcommand{\CMF}[1]{\mbox{$\bar{#1}(\lambda)$}}

\newcommand{\wdj}[3]{WD\,J#1$#2$#3}

\title[The diverse chromaticities of white dwarfs]{Any colour you like: The diverse chromaticities of white dwarfs}

\author[M. A. Hollands]{
Mark A. Hollands,$^{1}$\thanks{E-mail: Mark.Hollands@warwick.ac.uk}
Boris T. G\"ansicke,$^{1}$
Pier-Emmanuel Tremblay$^{1}$
and
Detlev Koester$^{2}$
\\
$^{1}$ Department of Physics, University of Warwick, Coventry CV4 7AL, UK \\
$^{2}$ Institut f\"ur Theoretische Physik und Astrophysik, University of Kiel, 24098 Kiel, Germany \\
}

\date{Accepted 2026 September 02; submitted in original form 2026 August 11}

\pubyear{\the\year{}}

\begin{document}
\label{firstpage}
\pagerange{\pageref{firstpage}--\pageref{lastpage}}
\maketitle

\begin{abstract}
Main sequence stars have perceived colours ranging from orange to white to blue
with increasing effective temperature. This is because their atmospheric
opacities cause only minor deviations from blackbody colours when integrated
over the broadband sensitivity curves of the human visual system. White dwarfs
can have much denser atmospheres, reaching $\sim 1$\,g\,cm$^{-3}$ for the
coolest known objects. If metals such as carbon, calcium, or sodium are present
in these ultra-dense atmospheres, extreme broadening of atomic and molecular
opacities can produce optical spectra that diverge significantly from
blackbodies. We investigated how the combination of temperature and chemical
composition can affect the perceived colours (chromaticity) of white dwarfs.
While we found that most white dwarfs are similar in colour to blackbodies, we
also demonstrate examples of known stars that would appear yellow, green, cyan,
pink, or purple if viewed close enough for their stellar discs to be resolved.
We examined the chromaticity distribution for the volume-limited 20\,pc local
sample of white dwarfs finding a median correlated colour temperature
of $\approx 6800$\,K, establishing that, on average, white dwarfs are indeed
white in colour. Finally we show how white dwarf spectral models can be used to
generate colour-accurate images of their stellar discs as they would appear from
up close.
\end{abstract}

\begin{keywords}
Data methods -- Visualisations -- Colour science -- White dwarfs
\end{keywords}



\section{Introduction}

Stellar objects exhibit a wide variety of spectral energy distributions,
depending on the physical properties of their outer layers, primarily including
the temperature, density, and chemical composition of the stellar atmosphere.
Consequently, these atmospheric conditions can influence how stars appear to a
human observer, with sunlight and Sun-like stars having a white appearance, and
hotter/cooler stars appearing pale blue/orange due to their approximate
resemblance to blackbodies \citep{ms_harre21}, though with the latter typically
referred to as red in scientific contexts. These colours are accessible without
the aid of telescopes, with bright stars such as Betelgeuse and Antares having
noticeably redder hues than other nearby objects on the sky, and those like
Sirius and Rigel appearing slightly bluer. However, the majority of stars are
not visible to the naked-eye, including faint stellar objects like M dwarfs,
brown dwarfs, and white dwarfs (the subject of this work). If these faint
sources are viewed through telescopes, their colours will still not be
accurately conveyed compared to if they were observed from up close, due to the
reduced colour-discrimination of the eye under dark conditions
\citep{tootell20}.

The chromatic diversity of faint stars can be approximately reflected in
multi-channel astronomical imaging, though these are typically not
colour-accurate. For example, images from the Sloan Digital Sky Survey
\citep[SDSS,][]{yorketal00-1, alametal15-1, blantonetal17-1} map the $i$, $r$,
and $g$ bands onto the red, green, and blue (RGB) colour channels, respectively.
However, the full range of human spectral sensitivity (see
Section~\ref{sec:colourtheory}) covers only the $g$ and $r$ bands, limiting the
colour-accuracy of such images.\footnotemark\ Such false-colour images are often
desirable for scientific analyses, with colours corresponding to wavelength
ranges covering specific spectral features (which may also fall outside of the
optical range entirely). While scientifically useful, false-colour images have
the potential to mislead as to the true appearance of stars and other
astrophysical bodies.

\footnotetext{For example, a spectrum dominated by H$\alpha$ emission will be
covered by the SDSS $r$-band, and hence mapped to the green RGB channel, despite
having a vivid red appearance in reality.}

Establishing the perceived colours of astrophysical objects has received
interest from the scientific community. Much focus has been given to main
sequence stars
\citep{ms_murdin81,ms_murdin87,ms_kohman05,ms_savino15,ms_harre21,ms_cardiel21}
demonstrating their approximately blackbody colours. Brown dwarfs
\citep{bd_burrows01, bd_burrows03, bd_cranmer21, bd_rering25} have also been
studied where their perceived colours are found to be magenta/violet, primarily
due to strong absorption from the Na-D doublet with highly broadened line wings
in the optical \citep{allardetal07-1}. Interest in perceived colours is not
limited to stellar objects, with \citet{neptune24} providing a recent
reprocessing of Voyager 2 images of Uranus and Neptune, finding the latter to be
a much less vivid shade of blue than previous images had shown, with both
planets now appearing similar shades of pale blue-green. The perceptual colours
of exoplanets have also been investigated \citep{planet_evans13}, with
\citet{planet_gao21} using global circulation models of planetary atmospheres
containing various aerosols, showing potentially stark differences in colour
between their day and night sides. However, to our knowledge no particular
attention has been paid towards the perceived colours of post-main sequence
objects such as white dwarfs, which is the focus of this work. 

White dwarfs\footnotemark\ are the dense stellar remnants left behind by the
majority of main sequence stars at the end of their lives. Their extreme
properties can result in atmospheres that are significantly more dense compared
to their main sequence counterparts, particularly for the oldest stars in the
white dwarf cooling sequence
\citep{kowalski14-1,blouinetal18-1,bergeronetal22-1}, resulting in broadband
opacities that can drastically alter their optical fluxes away from resembling a
blackbody spectrum and providing ongoing challenges in their modelling
\citep{saumonetal22-1}. It therefore stands to reason that the perceived colours
of white dwarfs may also depart significantly from the blue, white, and orange
exhibited by other kinds of stellar objects.

\footnotetext{Originally referred to as \emph{faint white stars} \citep{luyten22-1}
owing to the warm nature of the first few known objects, the term \emph{white dwarf}
was (to the best of our knowledge) coined by William Luyten in the Lick Observatory 
Bulletin XI.}

In this work we investigate the perceived colours of white dwarfs across a wide
range of temperatures, compositions, and their resulting spectral types. In
Section~\ref{sec:colourtheory}, we provide an introduction to the field of
colour science, and describe the techniques required to convert astrophysical
spectra to displayable colours on a typical computer monitor. In
Section~\ref{sec:wds} we apply these techniques to a variety of observed and
synthetic white dwarf spectra across the full range of white dwarf spectral
subtypes, demonstrating their diverse range of perceived colours. We also apply
these techniques to the volume-limited 20\,pc local sample, showing the typical
distribution of white dwarf colours. In Section~\ref{sec:vis}, we show how
colour-accurate images of white dwarfs can be generated from spectral models,
where we provide examples for a variety of typical and atypical white dwarfs.
Finally we present our conclusions in Section~\ref{sec:conc}.

\section{Calculating colours from spectra}
\label{sec:colourtheory}

While no stars other than the Sun can be observed close enough for their stellar
discs to be visually resolved, the information required to determine their
perceived colours is contained entirely within their optical spectra.
Replicating these perceived colours from astrophysical spectra requires a model
of the human visual system response to light (as a function of wavelength), a
quantitative description of colour, and an understanding of how image and
lighting technologies reproduce specified colours. These requirements are
addressed by the field of colour science, which has become well established over
the last century \citep{cie1931, Judd1975, Wyszecki1982}, particularly in its
application to display technology \citep{Foley1982, Hall1988}. In this section
we give a basic introduction to the concepts used in the field of colour science
which allow the conversion of a spectrum to the corresponding display colours.

Throughout this work, we work in the standard RGB colour space 
\citep[sRGB,][]{sRGB}\footnote{
For an optimal viewing experience, computer monitors should be set
to sRGB mode if available, and software used to lower the colour temperature for
reducing eye strain should be disabled, or colours will not appear as intended.
Ideally self-emissive displays should be used, as projected images will have
their colours mixed with ambient light.}
using the \texttt{python} package \textsc{colorpy}.\footnotemark\
This module contains routines for converting spectra into various colour spaces,
conversions between colour spaces, and translation into display values to be
used for computer displays.

\footnotetext{\url{http://www.markkness.net/colorpy/ColorPy.html} \\\indent
\url{https://github.com/markkness/ColorPy}}

\subsection{Converting spectra to the XYZ colour space}

The CIE (International Commission of Illumination/Commission Internationale de
l'\'Eclairage) 1931 XYZ colour space \citep{cie1931} is widely used in colour
science and industry to quantify the subjective experience of colour, modelling
the perception of colour in a linear additive way. Essentially this reduces the
complexity of a spectrum down to three tristimulus values $X$, $Y$, and $Z$, by
performing integrals \citep{Wyszecki1982} of the form,
\begin{equation}
    X = \int_{360}^{830} F_\lambda(\lambda) \cdot\CMF{x}\,\mathrm{d}\lambda ,
    \label{eq:ciexyz}
\end{equation}
and symmetrically for $Y/\CMF{y}$ and $Z/\CMF{z}$, where $\lambda$ is wavelength
in nm, $F_\lambda$ is flux density per unit wavelength, and \CMF{x}, \CMF{y},
and \CMF{z} are the CIE 1931 colour-matching functions.\footnotemark\ Evidently,
these integrals are analogous to calculating synthetic photometry of a spectrum
within some bandpass in an astronomy context.

\footnotetext{Corrections are available to the 1931 colour-matching functions
\citep{judd51,vos78}, and with more recently determined functions available from
the CIE \citep{stockman+sharpe2000,cie2006,cie2015}. However, CIE 1931 remains
ubiquitous for calibration of display technology, with the transformations to
the sRGB colour space also specified in reference to CIE 1931. We
therefore use the 1931 colour-matching functions for internal consistency.}

\begin{figure}
  \centering
  \includegraphics[width=\columnwidth]{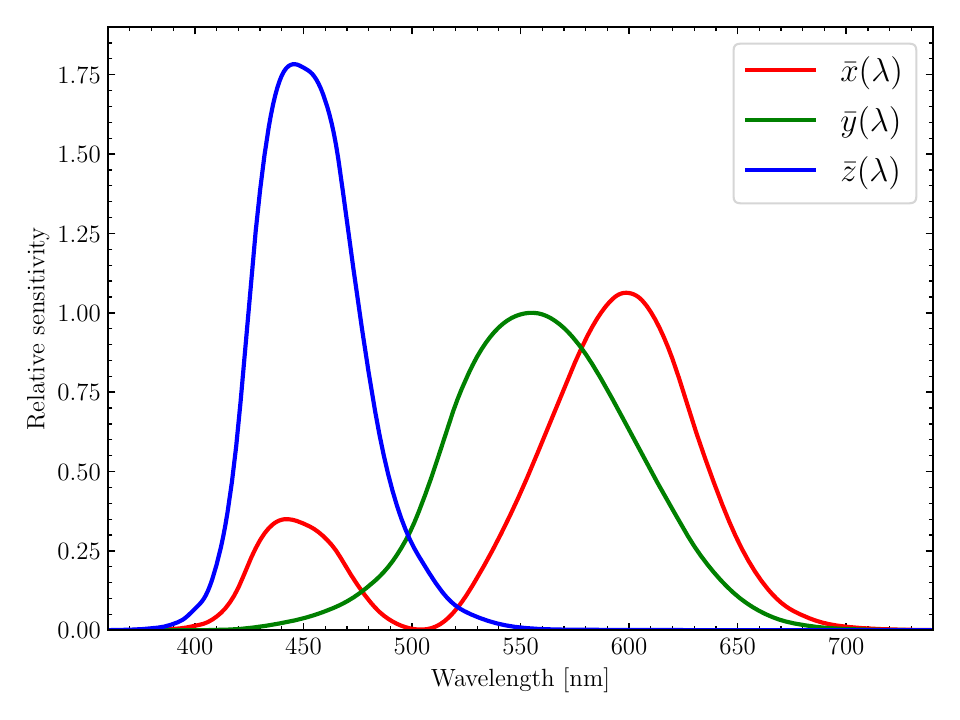}
  \caption{\label{fig:xyz_curves}
    Colour matching functions for the 1931 CIE XYZ colour space. 
    Integrating a spectrum over these curves results in the values
    of $X$, $Y$, and $Z$, which define the colour and intensity
    in the XYZ colour space.
    }
\end{figure}

The XYZ colour-matching functions (Fig.\,\ref{fig:xyz_curves}), were
empirically determined to reflect the human visual system response to light
under photopic vision over a 2\,degree field of view from the eye centre
\citep{cie1931}. It is important to note, however, that they do not directly
reflect the individual cone responses, which are instead covered by the LMS
colour space \citep[Long/Medium/Short, e.g.][]{stockman+sharpe2000}. Instead the
CIE colour-matching functions are defined such that all three are non-negative,
a flat spectrum in $F_\lambda$ yields equal values of $X$, $Y$, and $Z$, and
\CMF{y} is equivalent to the CIE photopic luminous efficiency function (the overall
human visual response to perceived light intensity as a function of wavelength
under photopic conditions) normalised to a peak of $1$ \citep{Wyszecki1982}.

In \textsc{colorpy} the function \verb+xyz_from_spectrum+ can be used on a
spectrum to calculate $XYZ$ values. Internally the integrals are performed using
straight sums, assuming constant wavelength separation between flux values. Spectra must
therefore be re-binned onto a linear wavelength-axis before this function
is used.

\begin{figure}
  \centering
  \includegraphics[width=\columnwidth]{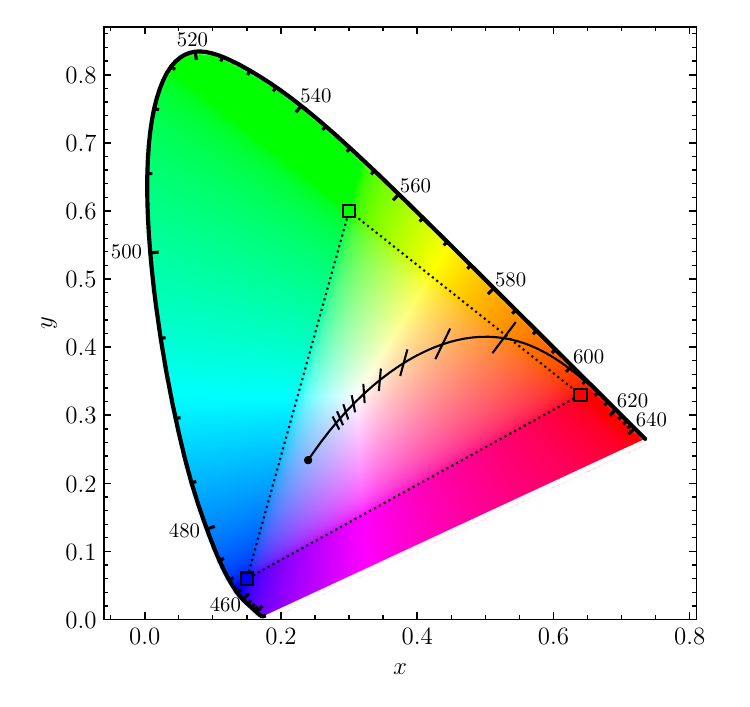}
  \caption{\label{fig:ciexy}
    CIE 1931 chromaticity diagram in the $x$-$y$ plane. The black curve bounding
    the filled region corresponds to monochromatic colours. The sRGB primaries
    are shown by the squares, defining a triangular region (dotted) which bounds
    the gamut of displayable chromaticities in the sRGB colour space. Colours
    outside of the triangle are therefore approximate. The black curve emanating
    from long wavelengths is the blackbody locus as a function of temperature,
    with isotherm ticks in correlated colour temperature shown from 2000\,K to
    10\,000\,K in steps of 1000\,K (from right to left).
    }
\end{figure}

While units of $F_\lambda$ are not specified above, they are mostly unimportant
here as all real stars have perceptual surface brightnesses (luminance, see
appendix \ref{sec:lum}) far beyond the maximum output of any computer monitor.
Instead we consider the component of colour independent of brightness, the
chromaticity. To determine the chromaticity \citep{Wyszecki1982}, $X/Y/Z$ are
normalised according to,
\begin{equation}
    x = \frac{X}{X+Y+Z},\qquad y = \frac{Y}{X+Y+Z},\qquad z = \frac{Z}{X+Y+Z}.
    \label{eq:xyz_norm}
\end{equation}
As there are now only two degrees-of-freedom remaining, $x$ and $y$ can be used
to define a 2-dimensional quantity that is independent of brightness, and is the
primary coordinate system used to describe chromaticity. We show the full CIE
$x$-$y$ chromaticity diagram in Fig.\,\ref{fig:ciexy}. The chromaticity is analogous
to a colour-colour diagram in astronomy contexts. However, rather than being
defined in terms of logarithmic flux ratios, normalisation by the total
restricts the $x$, $y$ values to the unit simplex.

The black curve around the edge of the filled region corresponds to
monochromatic colours. Since all physical spectra are a linear combination of
monochromatic colours, this curve bounds the space of all perceivable
chromaticities. Outside of this region (but within the unit simplex) the ratios of
$X/Y/Z$ are impossible given the CIE curves, for instance at long wavelengths
$>700$\,nm, the ratio $\CMF{x}/\CMF{y}$ converges to about 2.8, and so there are
no shades of red that involve illumination purely in $X$, but not in $Y$.
Such impossible colours can only be experienced via direct stimulation of
retinal cone cells \citep{olo_clr}.

The dotted triangle in Fig.~\ref{fig:ciexy} corresponds to the sRGB gamut, where
the corners (filled squares) correspond to the sRGB primaries (see
Section~\ref{sec:rgb}). Since this description of colour is linear, mixtures of
the sRGB primaries can only correctly reproduce colours interior to the
triangle. Consequently, chromaticities outside of the sRGB gamut cannot be
accurately represented. For display purposes, we approximate these by clipping
negative RGB values to zero.

The black curve emanating from the red corner is the blackbody/Planckian locus,
and corresponds to the chromaticities of blackbody spectra. Ticks are from
2000\,K to 10\,000\,K in 1000\,K steps. These ticks are not perpendicular to the
Planckian locus in the $xy$-plane, as they instead correspond to isotherms in
correlated colour temperature (see below). The 2000\,K step appears at the
boundary of the sRGB region, and so blackbody colours below that value cannot be
accurately represented by an sRGB monitor, though all blackbodies above
$2000$\,K can be. As no white dwarfs have convincingly been shown to have cooled
to effective temperatures (\Teff) below $3000$\,K
\citep{bergeronetal22-1,elmsetal22-1,caronetal23-1}, we expect most, if not all
white dwarf chromaticities to fall within the sRGB gamut. The black circle at
the end of the blackbody locus corresponds to $T\rightarrow\infty$, i.e. a pure
Rayleigh-Jeans tail where $F_\lambda(\lambda) \propto \lambda^{-4}$. This gives
$T\rightarrow\infty$ a chromaticity close to sky-blue, reflecting the
$\lambda^{-4}$ dependence of the Rayleigh-scattering cross-section.

While the CIE $x$-$y$ chromaticity diagram remains widely used, a limitation of this
colour space is that it is not perceptually uniform. In other words a small step
of, say $0.01$, within the $x$-$y$ plane will result in a perceived chromaticity
change that depends on the location within the plane and the direction of the
step. To address this non-uniformity, other colour spaces with improved
perceived uniformity have been introduced such as the 1960 Uniform Colour Space
(UCS) \citep{ucs1960macadam,ucs1960}. The UCS chromaticity values, $u$ and $v$ can
be determined directly from the CIE 1931 $x,y$ values, where
\begin{equation}
    u = \frac{4x}{3-2x+12y},\quad v = \frac{6y}{3-2x+12y}.
    \label{eq:uv}
\end{equation}
This form of chromaticity is used to define the correlated colour temperature,
and is achieved by minimizing the Euclidean distance from a given point in the
$u$-$v$ chromaticity plane to the blackbody locus (the ticks in
Fig.~\ref{fig:ciexy} were determined this way). This allows chromaticities near
to the blackbody locus to be assigned a temperature, which we make use of in
Section~\ref{sec:wds}. Note that the correlated colour temperature is
fundamentally different to the $\Teff$, which is instead derived from the entire
spectral energy distribution, though the two quantities may have similar values
for spectra resembling blackbodies.

\subsection{Conversion to linear RGB}
\label{sec:rgb}

A display colour space requires several key ingredients: the chromaticities of
three display primaries, the chromaticity of a white point: that is to say the
resulting mixture when all three primaries are at their maximum outputs, and a
set of transfer functions which map the linear intensities of the primaries to
display values.

In the sRGB colour space, the three primaries are labelled $R$, $G$, and $B$ and
have specified chromaticities $(x, y) = (0.64, 0.33)$, $(0.30, 0.60)$, $(0.15, 0.06)$,
respectively \citep{sRGB}. Their linear intensities are given on a dimensionless
scale 0 to 1, where 1 represents their maximum outputs. The white
point is given by the standard illuminant D65 (close to a 6500\,K blackbody
spectrum), with a specified chromaticity $(x, y) = (0.3127, 0.3290)$ \citep{sRGB}.
Finally, for an output of full-white, i.e. $R=G=B=1$, the luminance, $Y$, is
also specified to be 1. Collectively, these requirements can be used to derive a
transformation matrix from XYZ to sRGB (and vice versa),
\begin{equation}
    \left(\begin{array}{c}
    R \\
    G \\
    B \\
    \end{array}\right) =
    \left(\begin{array}{rrr}
     3.2410 & -1.5374 & -0.4986 \\
    -0.9692 &  1.8760 &  0.0416 \\
     0.0556 & -0.2040 &  1.0570 \\
    \end{array}\right) \cdot
    \left(\begin{array}{c}
    X \\
    Y \\
    Z \\
    \end{array}\right).
    \label{eq:rgb_from_xyz}
\end{equation}

Even if the XYZ colours are normalised as per equation\,\eqref{eq:xyz_norm},
after converting to sRGB (equation~\eqref{eq:rgb_from_xyz}) the linear RGB
values may not be at maximum intensity. Since the purpose of this work is to
demonstrate the appearance of stars, we wish for colours to be displayed as
brightly as possible, while preserving chromaticity. Therefore, throughout this
work, we always rescale sRGB colours so that the brightest RGB channel has a
linear value of $1$. In order to convert these linear values to display colours,
the final ingredient of the sRGB space, a transfer function must be applied.

\subsection{Transfer functions}
\label{sec:gamma}
In the sRGB colour space, 8-bit displayable values (per colour channel) are
given as integers in the range 0 to 255. One might na\"ively assume that
conversion from linear RGB values is as simple as,
\begin{equation}
    I_\mathrm{display} = \mathrm{round}( 255 I_\mathrm{linear} ),
    \label{eq:round}
\end{equation}
where $I$ are one of the RGB channels. However, the human eye does not respond
linearly to light intensity, and neither do pixel intensities as a function of
the 0--255 display values. Instead, an opto-electronic transfer function
(OETF) is used to map linear light values to a displayable range, and an
electro-optical function (EOTF) is used to perform the inverse. Historically,
cathode-ray tube displays had intensities with a power-law (gamma-function)
dependence on the voltage $V$,
\begin{equation}
    I \propto V^\gamma,
    \label{eq:gamma_curve}
\end{equation}
where typically $\gamma=2.2$. Instead of a pure gamma function, sRGB uses a
piecewise function \citep{sRGB}, which approximates a gamma function with
$\gamma=2.2$, but has a small linear segment for low intensities, 
avoiding a vertical slope at the origin as with a pure gamma OETF. For input
intensities $I$ and output values $V$, both in the range 0--1, the OETF is given
by
\begin{equation}
    V = \left\{
    \begin{array}{ll}
    12.92I & \mathrm{if}\quad I \leq 0.0031308 \\
    1.055I^{1/2.4} - 0.055 & \mathrm{otherwise} \\
    \end{array}
    \right. ,
    \label{eq:gamma_srgb}
\end{equation}
and is applied to all three RGB channels. After eq.\,(\ref{eq:gamma_srgb}) has
been applied to the linear values, then eq.\,(\ref{eq:round}) is used to convert
to the display values in the range 0--255.

\begin{figure*}
  \centering
  \includegraphics[width=\textwidth]{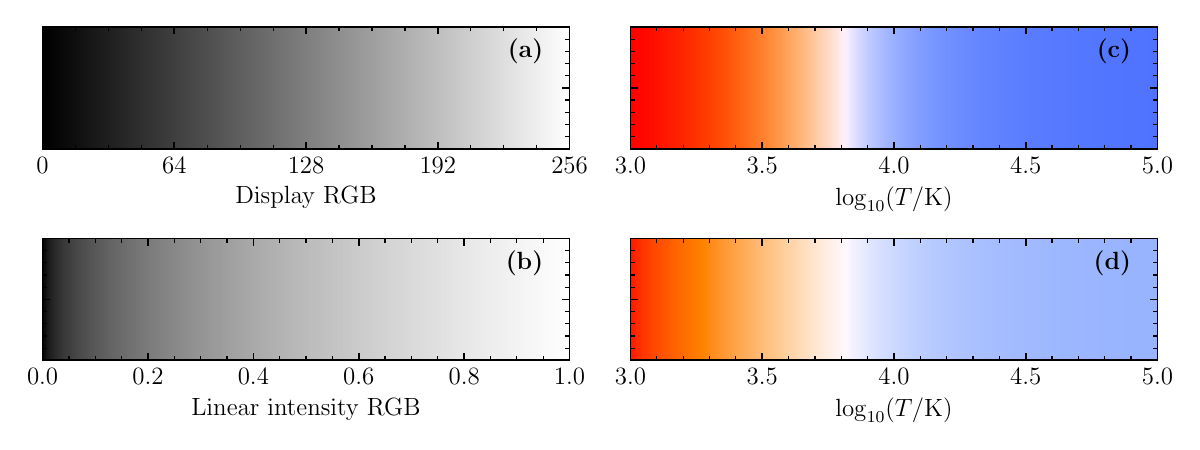}
  \caption{\label{fig:gamma}
  Comparison between not using (top) and applying (bottom) the sRGB OETF for
  grey-scale (left) and blackbodies (right). (a) The grey-scale is
  approximately perceptually linear, but due to the non-linear response of the
  human visual system, it is non-linear in intensity. (b) The
  grey-scale is linear in intensity, but appears perceptually non-linear.
  Blackbody colours appear overly saturated when an OETF is not applied (c)
  compared with applying the sRGB OETF (d). Note that temperatures below $\simeq
  2000$\,K ($\log_{10}(T/\mathrm{K})\simeq3.3$) are outside of the sRGB gamut
  and so are approximated. All colours are scaled to the maximum displayable
  brightnesses.
  }
\end{figure*}

We now demonstrate visually the difference applying the sRGB OETF makes when used or
not (Fig.\,\ref{fig:gamma}). In the panels (a) and (b), we show simple
grey-scales. In panel (a), equation~\eqref{eq:gamma_srgb} is not used and each
value along the $x$-axis corresponds to one of the display values in the range
0--255. This scale appears perceptually close to linear, despite the non-linear
scale in brightness. This is of course useful for displaying shades with a large
dynamic range in intensity. In panel (b), the intensity is instead linear on the
range 0--1, but appears perceptually non-linear with a rapid perceived change at
low values, demonstrating the human eye's improved discrimination between
dark shades compared to bright ones.

In the right panels of Fig.\,\ref{fig:gamma}, we show blackbody colours
without (c) and with (d) the sRGB OETF (equation~\eqref{eq:gamma_srgb}) applied
before quantising with equation~\eqref{eq:round}. In both panels colours are
normalised so that the brightest of the three RGB channels is at its maximum
(1.0 on the linear scale and 255 on the display scale). The temperature scale
ranges from 1000--100\,000\,K (logarithmically) showing the expected transitions
through red, orange, white, and blue colours. In (c), the uncorrected case
(equivalent to applying a gamma OETF with $\gamma=1$), the cool temperatures are
much redder than real blackbodies, and at hot temperatures are much bluer than
expected, as the outputs between the channels do not have the correct ratios.
These oversaturated colours resemble those found by \citet{ms_harre21} who did
not apply an OETF in their analysis of main sequence stars.

\subsection{Examples}

To summarise, the process to go from a stellar spectrum to an sRGB colour is as
follows.
\begin{itemize}
    \item Integrate the spectrum over the CIE XYZ colour-matching functions
    using equation~\eqref{eq:ciexyz} to determine the values of $X$, $Y$, and
    $Z$. If needed, normalise using equation~\eqref{eq:xyz_norm} to determine
    the chromaticity in terms of $x$ and $y$.
    \item Use the transformation matrix in equation~\eqref{eq:rgb_from_xyz} to
    convert XYZ values to linear sRGB in terms of $R$, $G$, and $B$, normalising
    the brightest component to $1$.
    \item Apply the sRGB OETF from equation~\eqref{eq:gamma_srgb} and quantise
    with equation~\eqref{eq:round} to convert to displayable sRGB values in the
    0--255 range.
\end{itemize}

\begin{figure}
  \centering
  \includegraphics[width=\columnwidth]{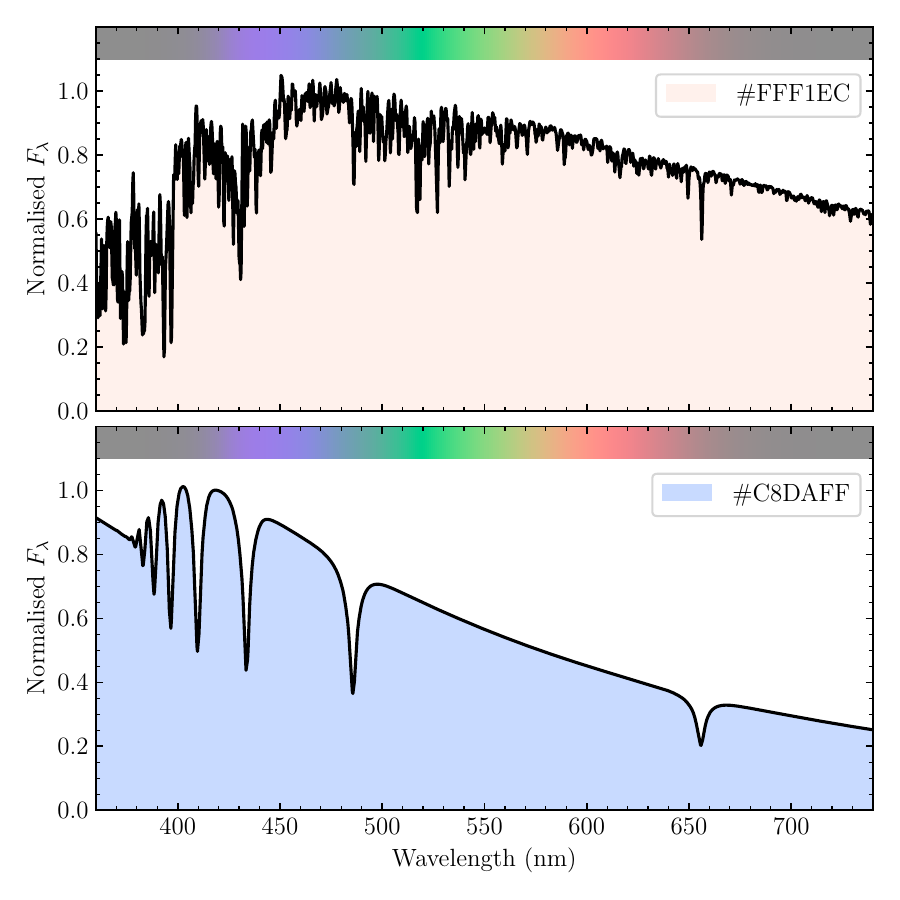}
  \caption{\label{fig:examplespec}
  A solar model spectrum (top) and a 10\,000\,K hydrogen atmosphere white dwarf
  model spectrum (bottom) are shown, under-filled with their perceived colours.
  Their corresponding sRGB hex codes are shown in the top right of each panel.
  Both models are convolved to a full width at half maximum of 5\,\AA\ for
  clarity. The bars above each panel approximate the appearance of monochromatic
  emission at constant intensity, though mixed with the  minimal amount of grey
  to move all colours in-gamut.
  }
\end{figure}

With these necessary ingredients specified above, we conclude this section with
two examples for a Solar model, and a pure hydrogen 10\,000\,K white dwarf model
spectrum, with the results shown in Fig.\,\ref{fig:examplespec}.

The Solar model in the top panel of Fig.\,\ref{fig:examplespec} was taken from
the BT-NextGen (AGSS2009) grid \citep{hauschildtetal99-1} with $\Teff=5800$\,K,
$\log g=4.5$, and solar metallicity. Integrating over the colour-matching
functions, and normalising to $Y=1$, we found ($X=0.974$, $Y=1$, $Z=1.024$),
with a corresponding chromaticity of $(x=0.325, y=0.334)$. Converting to linear
sRGB, and scaling to the brightest component, we arrive at $(R=1, G=0.878,
B=0.840)$. Applying the sRGB transfer function and quantising, we found display
values of $(R=255, G=241, B=236)$ (the corresponding hexadecimal value is shown
on the figure). While the Sun is ostensibly ``white'' in colour, the model
correlated colour temperature ($\approx 5870$\,K, slightly higher than its \Teff) is
somewhat lower than the 6500\,K sRGB white point, and so appears marginally
redder and fainter than the figure background.

The white dwarf model is of a pure hydrogen atmosphere with $\Teff=10\,000$\,K
at $\log g=8$ (see Section~\ref{sec:DADBs} for details). Following the same
procedure, we found ($X=0.960$, $Y=1$, $Z=1.496$) in XYZ space, with a
chromaticity of $(x=0.278, y=0.289)$. The normalised linear sRGB values were found
to be $(R=0.580, G=0.704, B=1)$. Converting to display sRGB, we then found
display values of $(R=200, G=218, B=255)$. Compared to the Solar model, the
$X/Y$ ratios are quite similar as they both approximately sample their
Rayleigh-Jeans tails. However the $Z/Y$ ratio is much higher for the white dwarf
due to the greater proportion of flux radiated in the blue end of the optical.
Consequently, the white dwarf appears pale blue, as expected for stars in the
region of $\Teff=10\,000$\,K.

Of course the surfaces of both example stars are enormously brighter than it is
possible to show here, but with the chromaticity of these stars depicted as
accurately as possible. As these model spectra represent the flux integrated
over their stellar discs, they can instead be interpreted as the colour they
would illuminate a white surface some distance from the stars, i.e. a surface
that scatters all wavelengths equally.

\section{Colours of white dwarfs}
\label{sec:wds}

\subsection{Pure hydrogen and helium atmospheres}
\label{sec:DADBs}

The simplest white dwarfs are those with near pure hydrogen or helium
atmospheres. For pure hydrogen atmospheres, their optical spectra are dominated
by hydrogen lines above $\Teff \simeq 5000$\,K giving them the spectral type DA
\citep{sionetal83-1}. Similarly, pure helium atmosphere white dwarfs exhibit
lines from neutral helium from around $11\,000$ to $40\,000$\,K giving them the
spectral type DB (above this range \Ion{He}{ii} lines dominate with spectra
designated DO). At lower temperatures, both hydrogen and helium are
insufficiently excited to show spectral lines, resulting in continuum spectra
classified as DCs.

To visualise the colours of these simpler white dwarfs, we used grids of models
with surface gravity, $\log g$, fixed to 8, and \Teff\ as the only free
parameter. Model spectra such as these are free from problems such as poor
flux-calibration which could make observed data yield incorrect colours. We made
use of the Tremblay model atmospheres, which include 3D and non-local
thermodynamic equilibrium (NLTE) effects
\citep{tremblayetal11-2,tremblayetal13-1}, and include a recent correction to
the H$_3^+$ partition function relating to nuclear spin degeneracy. More
recently these models have been adapted to compute 3D helium-rich atmospheres
\citep{cukanovaite18-1,cukanovaite19-1,cukanovaite21-1}. Here we focus on their
pure helium grid.

\begin{figure}
  \centering
  \includegraphics[width=\columnwidth]{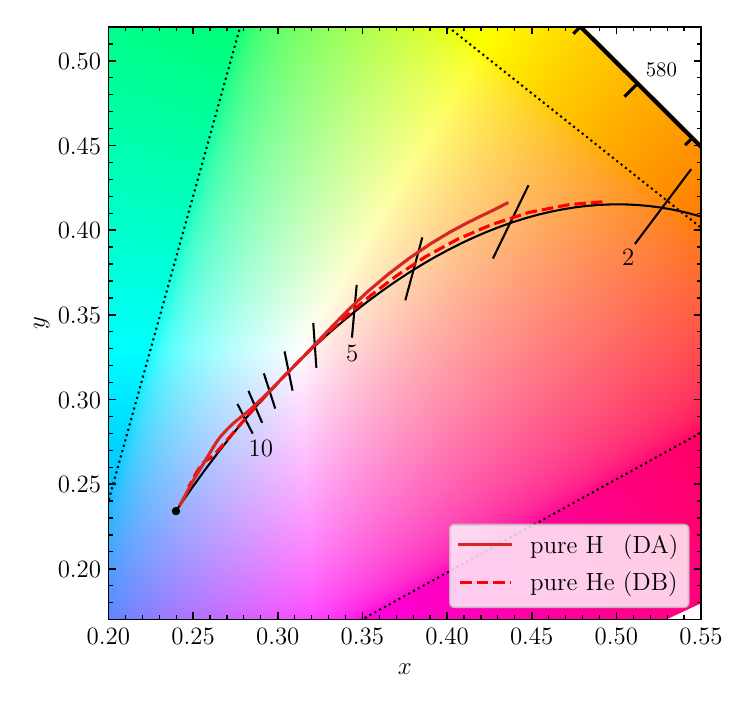}
  \caption{\label{fig:cieDADB}
    CIE chromaticity diagram, zoomed-in compared with Fig.\,\ref{eq:ciexyz}, and
    overlaid with tracks for pure hydrogen (DA) and pure helium (DB) white dwarf
    model colours as a function of \Teff, all with $\log g=8$. The black curve
    corresponds to the black body locus as before, with ticks indicating correlated colour
    temperature and their labels in kK.}
\end{figure}

For hydrogen atmospheres, we computed colours for models spanning
$\Teff=3000$\,K to $140\,000$\,K with step-size increasing as a function of
$\Teff$. For helium atmospheres, the model grid instead spanned 3000\,K to
40\,000\,K. The model tracks for both pure hydrogen and pure helium are shown in
the CIE chromaticity diagram in Fig.\,\ref{fig:cieDADB}. The chromaticity, linear
sRGB, and display sRGB hex codes are tabulated in Tables\,\ref{tab:DA} and
\ref{tab:DB}, and figures for a subset of those spectra are shown in
Figs.\,\ref{fig:DAspec} and \ref{fig:DBspec}.

From the tracks in Fig.\,\ref{fig:cieDADB}, we see that in general both pure
hydrogen and helium atmospheres are well approximated by blackbody colours.
However DAs do show some small deviations at the highest and lowest
temperatures. For correlated colour temperatures above $\simeq
9000$\,K, there is a slight deviation to higher $y$ values which vanishes for
even higher colour temperatures. We found this is due to the emergence and
disappearance of the Balmer lines in the optical, particularly H$\gamma$. While
the Balmer jump similarly emerges and vanishes in this \Teff\ range, it has less
of an effect on the chromaticity due to the diminished sensitivity of the
human-eye below 400\,nm (Fig.\,\ref{fig:xyz_curves}).

At correlated colour temperatures below about 5000\,K the hydrogen models again begin to
deviate away slightly from the blackbody colours towards higher $y$ values.
This can be attributed to two reasons: firstly, the formation of hydrogen
molecules at low $\Teff$ where collisions between H$_2$ and other neutral
particles \citep[H$_2$-H$_2$ or H$_2$-H, ][]{borysowetal01-1} can induce a
temporary H$_2$ dipole moment, known as collision induced absorption (CIA).
Initially the broad opacity from CIA only has a measurable effect on the near
infrared, only subtly affecting optical fluxes. At even lower $\Teff$, the CIA
becomes sufficiently strong to be visible at optical wavelengths, where it is
apparent at the reddest wavelengths of the $\Teff=3000$\,K model in
Fig.~\ref{fig:DAspec}. Secondly, neutral broadening of the hydrogen Ly$\alpha$
line in the ultraviolet, affects the blue end of the optical at low \Teff\
\citep{kowalski+saumon06-1}. The broadening of Ly$\alpha$ arises due to
interaction with other particles (protons, electrons, H, H$_2$) in the plasma
(which becomes increasing dense at low \Teff), altering both the hydrogen energy
levels and dipole moments. At the lowest \Teff, Ly$\alpha$ broadening becomes so
extreme that its red wing extends into the blue end of the optical, depressing
flux in this region \citep{rohrmannetal11-1}.

By contrast, the pure helium track closely follows the blackbody locus in
Fig.~\ref{fig:cieDADB}. At low \Teff\ the optical opacities are almost grey,
arising predominantly from He$^-$ free-free absorption: due to the absence of
hydrogen in these models, no broad Ly$\alpha$ is present, nor the CIA opacities
mentioned above. Inspecting the models in Fig.~\ref{fig:DBspec} it is indeed
clear that the featureless spectra of pure helium atmospheres resemble
blackbodies \citep{suzuki+fukugita18-1,serenellietal19-1} for the $\Teff \leq
10\,000$\,K models. Another noteworthy point for our pure helium models is that
while they have chromaticities closely resembling blackbodies, their \Teff\
appears higher than the corresponding colour temperature.
This is most easily seen at the cool end of the helium track in
Fig.~\ref{fig:cieDADB}, with a corresponding $\Teff$ of $3000$\,K, but at a
correlated colour temperature of about $2400$\,K. For DAs the
difference is smaller and inverted: with the $\Teff=3000$\,K end of the track
having a colour temperature of about 3100\,K instead. A
final important point of interest is that both model tracks shown in
Fig.\,\ref{fig:cieDADB} are fully contained well within the sRGB gamut,
indicating that in general, white dwarfs chromaticities can be accurately
represented by sRGB displays.

\subsection{Cool, mixed atmosphere (IR-faint) white dwarfs}
\label{sec:UC}

In reality, white dwarfs with effectively pure helium atmospheres are rarely
encountered, with at least some level of trace hydrogen usually present. This
hydrogen provides an additional source of free electrons, and introduces other
opacity sources that are not present in pure helium. Since hydrogen is present,
the Ly$\alpha$ line is once again present in the ultraviolet, but now also
broadened by neutral helium \citep{gaensickeetal18-1,allardetal20-1}. Another
important opacity source becomes prevalent at low \Teff\ ($<5000$\,K), where
trace hydrogen introduces two new forms of CIA from H-He and H$_2$-He, the
latter of which can become particularly strong for abundances of $\logXY{H}{He}
\simeq -3$\,dex \citep{blouinetal17-1}. These CIA features can severely depress
flux in the near infrared earning them the classification IR-faint\footnotemark\
\citep{bergeronetal22-1}. In the most extreme cases the CIA can extend far into
the optical, resulting in white dwarfs with optical spectra peaking at blue
wavelengths, despite their low temperatures.

\footnotetext{Such stars had previously been referred to as ultra-cool white
dwarfs, a term reserved for $\Teff < 4000$\,K objects. However,
\citet{bergeronetal22-1} showed that corrections to H$_2$-He CIA and He$^{-}$ free-free
opacities results in higher measured \Teff.}

IR-faint white dwarfs are among the most challenging degenerate stars to model
in terms of their spectra \citep{bergeron+leggett02-1}, as their transparent
atmospheres reach liquid densities requiring advanced equation-of-state models
\citep{blouinetal24-1}. Even the most advanced white dwarf model atmosphere
codes struggle to fully model their ultra-broad spectral features, including
those we have access to. To avoid the limitations of our models causing
incorrect calculated colours, for IR-faint white dwarfs we have opted to use
optical spectra from SDSS.

To ensure problematic flux calibrations of SDSS spectra did not adversely
influence our colour calculations, we adopted the method from
\citet{hollandsetal17-1} for correcting SDSS spectral fluxes using their
photometry. The SDSS spectra fully cover the wavelength ranges for the $g$, $r$,
and $i$ bands from SDSS imaging. We therefore calculated synthetic photometry
from SDSS spectra for these three, minimizing the magnitude
differences by iteratively re-scaling the spectra by first-order polynomials in
magnitude-space (exponential in flux-space) as a function of wavelength. In
\citet{hollandsetal17-1}, we found that this approach worked well, where the
fitted models to DZ spectra agreed with the other SDSS photometric bands ($u$
and $z$) that are not fully covered by the SDSS spectra. In some cases
subtraction of \Ion{O}{i} and \Ion{Na}{i} sky-emission lines is poor due to the
faintness of some of the objects, and so we interpolated over the wavelengths
for those lines to remove their influence on our colour calculations. Additionally,
SDSS spectra taken before data release 9, when the SDSS spectrograph was upgraded
to the BOSS instrument \citep{dawsonetal13-1}, only extend as
far bluewards as 380\,nm. While the CIE colour-matching functions are tabulated
down to 360\,nm, by 380\,nm the \CMF{z}-curve (Fig.\,\ref{fig:xyz_curves})
sensitivity is already negligible, and only 0.05\,percent of its integrated area
lies between 360 to 380\,nm. This wavelength range is therefore ultimately
unimportant for calculating accurate chromaticities.

\begin{figure}
  \centering
  \includegraphics[width=\columnwidth]{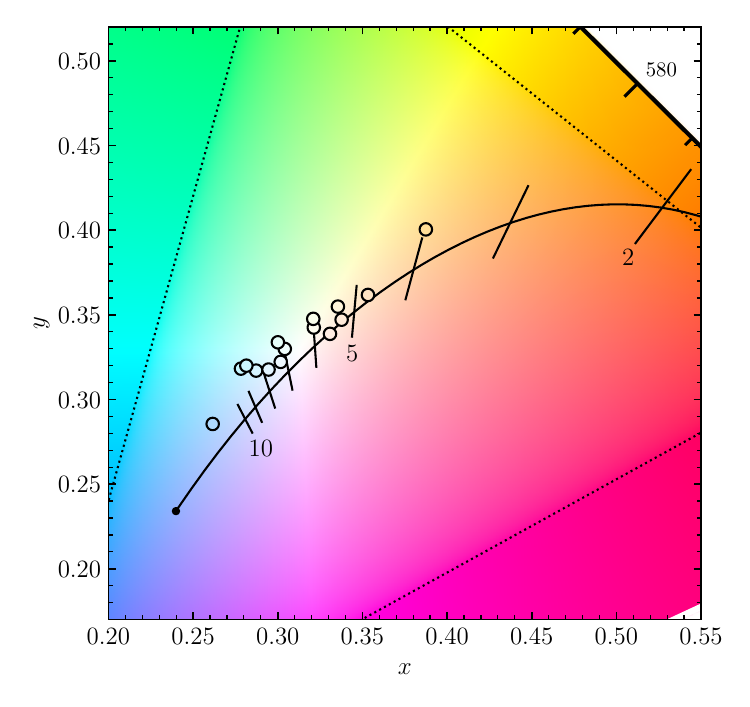}
  \caption{\label{fig:cieUC}
    CIE chromaticity diagram, zoomed-in compared with Fig.\,\ref{eq:ciexyz},
    showing a selection of IR-faint white dwarfs with optical flux-peaks shifted
    bluewards. The most extreme objects appear at cyan chromaticities.
    }
\end{figure}

We selected 15 IR-faint white dwarfs (Table\,\ref{tab:UC}) from
\citet{hollandsetal17-1} and \citet{bergeronetal22-1} with peculiar spectra
(Fig.\,\ref{fig:UCspec}) for which we determined their perceived colours and
locations in the CIE chromaticity diagram (Fig.~\ref{fig:cieUC}). These stars
were chosen to demonstrate the range of chromatic diversity for objects with
extreme CIA affecting their optical colours, and therefore should not be
considered an unbiased or representative sample of IR-faint white dwarfs in
general. Despite being ostensibly cool stars, most objects considered have
correlated colour temperatures between 5000\,K and 10\,000\,K,
demonstrating the strong effect that the CIA opacity has on their emergent flux
(compared to pure helium) and thus the chromaticity. However, this does not mean
that IR-faint stars simply have colours resembling warmer blackbodies, instead
having chromaticities shifted upwards in the $y$-direction compared to the
blackbody locus. From Fig.~\ref{fig:cieUC} and \ref{fig:UCspec} it is seen that
the IR-faint white dwarfs with the lowest $x$ values occupy the cyan part of the
chromaticity diagram. In Fig.\,\ref{fig:UCspec}, this is perhaps most readily
apparent for \wdj{0039}{+}{3035} and \wdj{1403}{+}{4533} (full coordinates given
Table~\ref{tab:UC}), which both show relatively high signal-to-noise ratio (S/N)
spectra and prominent cyan colours. Even more extreme is \wdj{1251}{+}{4403}
which appears visually blue with a correlated colour temperature of 12\,700\,K,
though we note its low S/N spectrum. The IR-faint white dwarfs with
correlated colour temperatures near 5000\,K appear a pale yellow
colour instead. By contrast \wdj{0854}{+}{3503} shows relatively little CIA in
the optical, and therefore has an orange colour more befitting its \Teff. In any
case, IR-faint white dwarfs demonstrate that the broad opacities encountered in
these ultra-dense atmospheres result in stars that are far from blackbodies at
the same $\Teff$.

\subsection{Carbon rich white dwarfs}
\label{sec:DQs}

White dwarfs with spectra dominated by carbon features are classified as DQs
\citep{sionetal83-1}. The carbon is generally understood to be dredged up core
material, mixed into a convective outer helium layer
\citep{koesteretal82-2,fontaineetal84-1}. For hot DQs a merger origin is also
possible \citep{dufouretal07-1} resulting in carbon-dominated atmospheres. The
most common case, and the one we predominantly examine here, is for cool white
dwarfs where helium features have practically vanished and the dominant carbon
features are from C$_2$ molecules. These molecules form wide absorption bands
across the optical known as the Swan bands, and are also visible in carbon stars
\citep{bernath09-1}. As these bands occur at wavelengths the human eye is most
sensitive to, it stands to reason that the perceived colours of DQs could be
dramatically different from blackbodies.

For \Teff\ below $\simeq 6500$\,K where photospheric densities of helium rich
atmospheres exceed $\simeq0.15$\,g\,cm$^{-3}$ \citep{blouinetal19-4}, the Swan
bands become distorted and shifted bluewards
\citep{hall+maxwell08-1,kowalski10-1}. Further distortion may also result from
magnetic fields \citep{bues99-1}. These ``peculiar'' DQs are classified DQpec to
distinguish their unusual spectra \citep{schmidtetal95-2}. Such spectra pose a
modelling challenge to state-of-the-art white dwarf atmosphere codes, with the
most extreme examples only approximately replicated \citep{blouinetal19-4}. For
this reason we again opt to use observed spectra from SDSS, applying the same
correction to flux calibration as in Section~\ref{sec:UC}.

\begin{figure}
  \centering
  \includegraphics[width=\columnwidth]{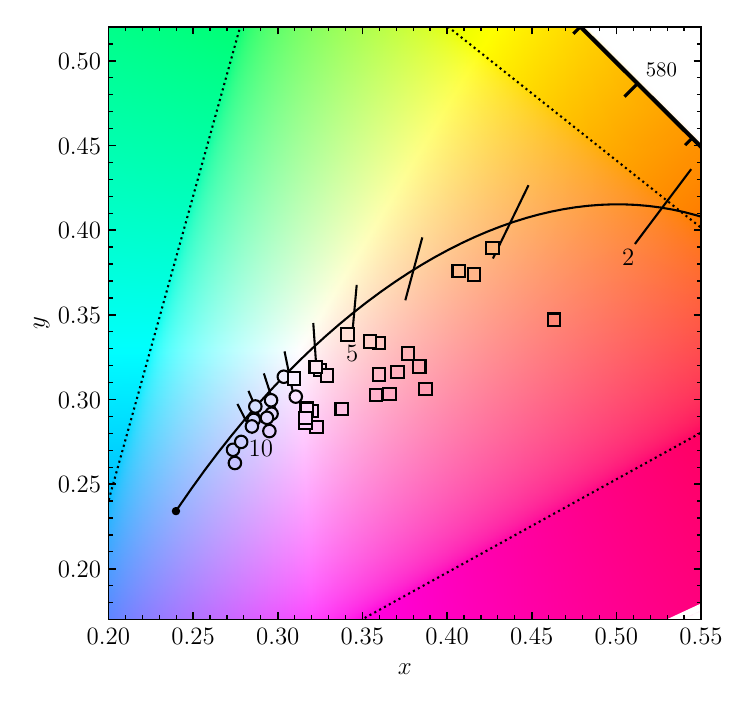}
 helium dominated  \caption{\label{fig:cieDQ}
    CIE chromaticity diagram, zoomed-in compared with Fig.\,\ref{eq:ciexyz},
    showing a selection of DQs, both normal (circles) and peculiar (squares) DQ subtypes.
    Generally the DQs appear below the blackbody-locus, giving them pink and purple colours.
    }
\end{figure}

We drew a selection of 10 DQpec white dwarfs from \citet{blouinetal19-4}, where
most of their objects of study have SDSS spectra. We also selected 26 more DQ
and DQpec white dwarfs with SDSS spectra \citep{kepleretal16-1} to complement
these, bringing the total to 36. As with IR-faint white dwarfs this is not meant
to be a representative, unbiased sample of DQs, but is intended to demonstrate
the wide range and most extreme white dwarf colours that are possible. As with
the other white dwarf classes we show the DQ colours in a zoomed-in chromaticity
diagram in Fig.\,\ref{fig:cieDQ}. The spectra themselves are shown in
Figs.\,\ref{fig:DQspec1} and \ref{fig:DQspec2}, and their colour information is
listed in Table\,\ref{tab:DQ}.

In contrast with the IR-faint white dwarfs the DQs are shifted below and to the
right of the blackbody locus in the chromaticity diagram. As with the previous
spectral classes, all DQ white dwarf colours are contained with the sRGB gamut
indicating they can be replicated accurately. The warmest of the selection are
shades of blue close to blackbodies, such as \wdj{0859}{+}{3257}
(Fig.\,\ref{fig:DQspec1}) which shows a mix of spectral features from atomic and
molecular carbon. For slightly lower correlated colour temperatures, a
few DQs appear as pale shades of purple instead, such as \wdj{0136}{+}{2004}.
Below colour temperatures of about 7000\,K, most objects
show minor to extreme distortion of the Swan bands (DQpec), and are instead
indicated by squares in Fig.\,\ref{fig:cieDQ}. These objects mostly exhibit
various shades of pink depending on their temperature and the strength of the
Swan bands (Figs~\ref{fig:DQspec1} and \ref{fig:DQspec2}). Some notable DQpec
include the magnetic objects \citep{schmidtetal03-1} \wdj{1113}{+}{0146} which
has highly distorted Swan bands, and \wdj{1333}{+}{0016} which is particularly
far from the blackbody locus $(x=0.387, y=0.306)$ giving it a more vivid pink
hue than other similar objects. At cool temperatures with lower carbon
abundances we also found DQpec with peach colours instead, such as
\wdj{1803}{+}{2320} (Fig.\,\ref{fig:DQspec2}).

A standout DQpec is \wdj{1011}{+}{2845} (also known as LHS\,2229 or
LP\,315$-$42). This nearby white dwarf \citep[$14.73\pm0.01$\,pc
away,][]{obrienetal24-1}, is estimated to have a low $\Teff$ of $4300$\,K
\citep{blouinetal19-4}, and has a spectrum (Fig.\,\ref{fig:DQspec2}) of deep and
distorted Swan bands, which likely arises from a combination of atmospheric
density \citep{blouinetal19-4} and a strong magnetic field
\citep{schmidtetal99-1}. As a result, the majority of its visible flux is
restricted to wavelengths $>600$\,nm, with wavelengths from $400$--$600$\,nm
greatly suppressed. The resulting colour is a unique hue of deep salmon-pink.
Like \wdj{1333}{+}{0016} (which has similarly shaped Swan bands, but emits a
greater fraction of flux below 450\,nm) it also resides far from the Planckian
locus in Fig.~\ref{fig:cieDQ}, making it one of the most visually distinct
objects in this analysis, and likely the reddest star known in terms of
perceived colour. White dwarfs such as \wdj{1011}{+}{2845} clearly demonstrate
the effect that broadband opacity sources such as molecular bands can have on
their atmospheres, making them visually different from the orange, white, and
blue chromaticities of main sequence stars and blackbodies.

\subsection{Cool metal rich white dwarfs}
\label{sec:DZs}

When metal lines (excluding carbon) appear in the spectra of white dwarfs a Z is
added to their spectral classification. These metals are now understood to
originate from the external pollution of the white dwarf atmosphere by rocky
debris from its planetary system, i.e. asteroids \citep{jura03-1,
zuckermanetal07-1, zuckermanetal10-1} and comets \citep{xuetal17-1,
williamsetal25-1, sahuetal25-1}, with almost 2000 such systems known at the
present time \citep{williamsetal24-1}. Warm white dwarfs enriched with planetary
traces are classified DAZ or DBZ when metal lines are secondary to hydrogen or
helium, respectively, though their narrow metallic absorption lines have little
effect on their perceived colours compared to metal-free DAs and DBs. However,
once helium-rich white dwarfs become sufficiently cool ($\Teff < 11\,000$\,K),
helium lines are no longer visible giving them the classification DZ. Their
atmospheres become increasing transparent and dense with decreasing \Teff\
(reaching up to 1\,g\,cm$^{-3}$), causing the strongest metal lines to become
extremely broad (hundreds to thousands of \AA), and in some cases asymmetric
\citep{allardetal16-1}. These highly broadened metal lines therefore have the
potential to alter their chromaticities away from blackbodies in a comparable way
to carbon in cool DQs (Section~\ref{sec:DQs}).

Such lines cannot be modelled under the impact approximation with standard
Lorentzian profiles \citep{allard+kielkopf82-1}, and must instead be given a
bespoke treatment using the unified line shape theory \citep{allardetal99-1}.
The unified theory considers the affect of perturber distance (in this case
helium) on both the energy levels and dipole moments of the atom/ion undergoing
a transition, allowing the line profile to be determined as a function of
temperature and perturber density. For DZs, unified profiles are employed for
the strongest lines where accurate modelling of the line core and wings are
both important, specifically \Ion{Mg}{ii} h+k ($\simeq 2800$\,\AA), the
\Ion{Mg}{i} resonance line ($\simeq 2853$\,\AA) and b-triplet ($\simeq 5170$\,\AA),
the \Ion{Ca}{ii} H+K lines ($\simeq 3950$\,\AA), the \Ion{Ca}{i} resonance line
($\simeq 4227$\,\AA), and the \Ion{Na}{i} D-doublet ($\simeq 5890$\,\AA).
Calculations for these \citep{allard+alekseev14-1, allardetal16-1,
allardetal16-2, hollandsetal17-1} are now present in white dwarf atmospheric
codes for modelling DZs \citep[e.g.][]{hollandsetal17-1, blouinetal18-1,
coutuetal19-1, kawkaetal21-1}, and are used to varying degrees in the DZ models
we show here.

\begin{figure}
  \centering
  \includegraphics[width=\columnwidth]{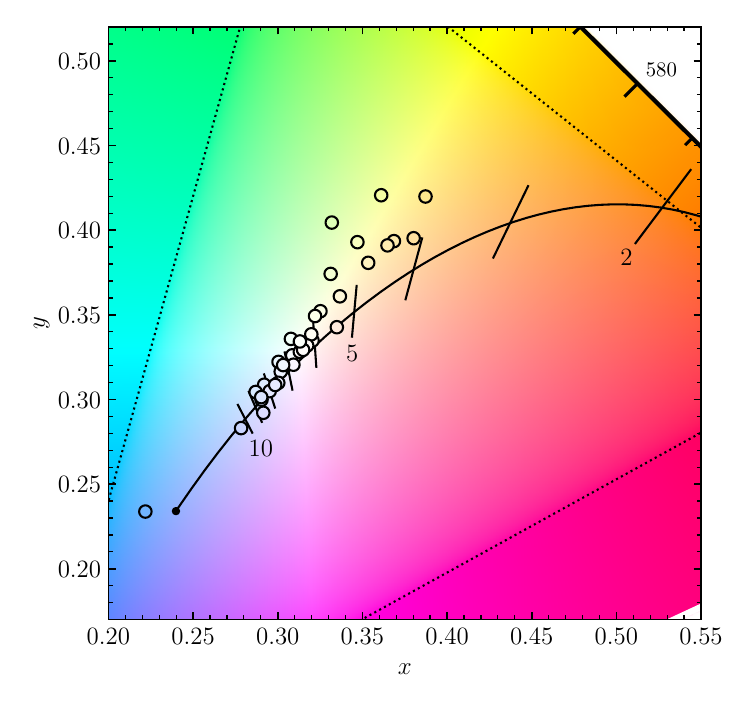}
  \caption{\label{fig:cieDZ}
    CIE chromaticity diagram, zoomed-in compared with Fig.\,\ref{eq:ciexyz},
    showing our sample of 36 DZs. Towards low temperatures most of the DZs
    appear above the blackbody locus, with \wdj{1330}{+}{6435} appearing
    green, and \wdj{2356}{-}{2054} a vivid blue.
    }
\end{figure}

We selected 33 DZ white dwarfs from \citet{hollandsetal17-1}, a sample of 231
DZs from the Sloan Digital Sky Survey (SDSS), for which we had already computed
model spectra using the 1D LTE Koester model atmosphere code
\citep{koester10-1}. Our 33 object selection was chosen to include examples of
typical DZs with strong metal lines across a wide range of temperatures ($4000 <
\Teff < 9000$\,K), as well as white dwarfs that we found had particularly
unusual colours. While there have been numerous improvements to the Koester
model atmosphere code since these models were first published, it is unnecessary
to refit the spectra: As long as the models provide reasonable matches to both
the optical spectra and photometric fluxes then the calculated chromaticities will
remain accurate. We therefore used updated models only where already available
from more recent analyses \citep[i.e. \wdj{0956}{+}{5912}, \wdj{1038}{-}{0036},
and \wdj{1535}{+}{1247},][]{hollandsetal22-1} or for objects where further
modelling was required in Section~\ref{sec:vis} (\wdj{0744}{+}{4649},
\wdj{0851}{+}{1543}, and \wdj{0916}{+}{2540}). We supplemented the above 33 DZs
with three more objects with particularly intriguing spectra:
\wdj{1330}{+}{6435} and \wdj{1824}{+}{1213} from \citet{hollandsetal21-1}, and
\wdj{2356}{-}{2054} from \citet{blouinetal19-1}, bringing the total selection to
36. These stars were also modelled with an up-to-date version of the Koester
atmosphere code. The 36 selected objects are shown in a chromaticity diagram
(Fig.\,\ref{fig:cieDZ}), with their points filled with their corresponding
colours. All objects are interior to the sRGB gamut indicating that the colours
of DZs can be displayed accurately, as with other white dwarf spectral classes.
Their spectra are shown in Figs.\,\ref{fig:DZspec1} and \ref{fig:DZspec2}, and
their colour information is listed in Table\,\ref{tab:DZ}.

There is a moderate degree of scatter in the model chromaticities at all
temperatures (Fig.\,\ref{fig:cieDZ}), though for the warmer objects with
correlated colour temperatures above $6000$\,K, this scatter is mostly
around the blackbody locus. At cooler temperatures the white dwarfs move towards
higher $y$-values where all objects with correlated colour
temperatures below $5000$\,K sit above the blackbody locus. Consequently, the
colours of these stars exhibit a particularly diverse range of hues. We discuss
some of the more uniquely interesting objects below.

\wdj{1330}{+}{3029} despite its strong metal lines has a results in a chromaticity
that almost exactly matches the sRGB white point, with a hex code of \#FFFEFF.
\wdj{0744}{+}{4649} shows strong absorption at blue wavelengths from pressure
broadened resonance lines of \Ion{Ca}{i} and \Ion{Ca}{ii}
(Fig.\,\ref{fig:DZspec1}). It also exhibits a deep and unresolved \Ion{Na}{i}
doublet at 589\,nm at orange wavelengths, but does not show much absorption from
Mg at 517\,nm (green wavelengths), as with most DZs. These abundances led
\citet{hollandsetal18-1} and \citet{harrisonetal21-1} to conclude the accreted
planetesimal contained a moderate amount of crust-like material. As a result of
these absorption features (particularly Ca which absorbs most of the flux below
460\,nm) its perceived colour is not orange, but yellow. Other white dwarfs such
as \wdj{0916}{+}{2540} (Fig.~\ref{fig:DZspec1}) are also seen to be yellower
than is possible for blackbodies, but still contain some degree of orange.

\wdj{1646}{+}{2238} (Fig.\,\ref{fig:DZspec2}) is markedly different from the
other DZs, having a larger than typical amount of trace hydrogen of $\logX{H} =
-1.8$, which results in its continuum slope being shallower than other DZs of
the same \Teff, increasing the degree of red flux. As a result, it has a
pale-pink colour, though not as vibrant as found for the DQs in
Section~\ref{sec:DQs}.

The most surprising colour comes from \wdj{1330}{+}{6435}. This object has been
analysed by \citet{kaiseretal21-1} and \citet{hollandsetal21-1}, due to the
presence of lithium in its spectrum (Fig.\,\ref{fig:cieDZ}), with another recent
analysis by \citet{blouinetal19-1}. For the analysis in
\citet{hollandsetal21-1}, we lacked data for unified profiles for the Na doublet
at that time, instead using inadequate Lorentzian profiles that led to
a cooler solution than \citet{kaiseretal21-1}. With unified profiles
for Na now in hand we find similar model parameters to
\citet{kaiseretal21-1} with $\Teff=4240$\,K, $\log g=8.22$\,dex,
$\logX{Li}=-10.6$, $\logX{Na}=-8.7$ and $\logX{Ca}=-8.8$. Due to the strong,
broad \Ion{Na}{i} doublet absorbing orange wavelengths, and Ca absorbing at
violet wavelengths, the resulting peak in flux occurs around 500\,nm, giving
\wdj{1330}{+}{6435} a perceived colour of chartreuse green
(Fig.\,\ref{fig:DZspec2}). To our knowledge, no other stars of any kind have
been claimed to exhibit a green hue, making white dwarfs so far the only known
way to produce a green star. We note that the spectrum of \wdj{1330}{+}{6435}
is somewhat noisy, and so it is conceivable that its chromaticity may differ
slightly in reality. However, our previous model from \citet{hollandsetal21-1}
using a standard Lorentzian profile shows a similar hue (\#EAFFA1),
giving us confidence in the green appearance of \wdj{1330}{+}{6435}.

\wdj{1330}{+}{6435} is not the only green object in our sample of 36, though it
is the most vivid: \wdj{0851}{+}{1543} (Fig.\,\ref{fig:DZspec1}), and
\wdj{1351}{+}{2645} (Fig.\,\ref{fig:DZspec2}) also both appear green but are
much paler than \wdj{1330}{+}{6435}, where their green hues can be explained by
strong \Ion{Ca}{ii} at blue wavelengths, and their steep Rayleigh-Jeans tails in
the red. For \wdj{0851}{+}{1543}, while its colour is pale, the hue is further
from yellow than the other objects (between sRGB green and cyan). Since the fit
to this star was based on a higher S/N spectrum, the green status of
\wdj{0851}{+}{1543} can be claimed reliably.

We also investigated \wdj{1644}{+}{1234}, another Li-rich white dwarf analysed
by \citet{kaiseretal21-1}. While we do not have access to their data, we were
able to compute a model with the same parameters. Our model showed a similar
greenish colour (\#EAFFC5) to \wdj{1330}{+}{6435}, but exhibited an excess of
flux between 4600--5100\,\AA\ not seen in the \citet{kaiseretal21-1} data or
model, indicating our model could not be used reliably, and so is not shown
here. Extracting the model from their fig.~1, we instead found a sandy
yellow colour (\#FFEAAB) more appropriate to its low \Teff.

The DZ white dwarf with a spectrum most severely departing from a
blackbody is \wdj{2356}{-}{2054} which was first properly analysed by
\citet{blouinetal19-1}. This star was originally discovered by
\citet{oppenheimeretal01-1}, with its spectrum initially described as
``bizarre'' due to ultra-broad \Ion{Na}{i} absorption spanning a large
proportion of the optical. The analysis of \citet{blouinetal19-1} showed that
this object could be accurately modelled after recently developed unified
profiles of Na-He were included. The spectrum also contains a molecular band
from MgH blended with a quasi-molecular satellite of Na-He. We calculated a
model with identical parameters to \citet{blouinetal19-1} which well reproduces 
their model. The extreme Na absorption is found to practically wipe out visible
emission in the centre of the optical, with the perceived colour primarily being
weighted to fluxes in the 450--500\,nm range. This results in a vivid blue
appearance (Fig.~\ref{fig:DZspec2}), despite having a $\Teff \simeq 4000$\,K. In
the chromaticity diagram (Fig.~\ref{fig:cieDZ}), \wdj{2356}{-}{2054} is found to
be bluer than even a pure Rayleigh-Jeans tail meaning this low \Teff\ star
cannot be assigned a colour temperature.

While \wdj{2356}{-}{2054} appears in a very different part of the chromaticity
diagram, it is qualitatively similar to \wdj{1330}{+}{6435} in that they are
both cool ($\Teff \simeq 4000$\,K) white dwarfs with helium dominated
atmospheres, and with \Ion{Na}{i} being the dominant spectral feature (also with
prominent absorption from \Ion{Ca}{i} at 4227\,\AA). Therefore
\wdj{2356}{-}{2054} essentially represents a more extreme version of
\wdj{1330}{+}{6435} where, given a large enough sample of cool DZ white dwarfs,
a continuum of objects likely exists between them allowing for other DZs with
vivid green, cyan, and blue hues which may be identified in future surveys.

\subsection{\texorpdfstring{The 20\,pc local sample}{The 20 pc local sample}}
\label{sec:20pc}

We conclude Section\,\ref{sec:wds} by examining the chromaticities of all white
dwarfs within 20\,pc of the Sun. This unbiased volume-limited sample contains
152 white dwarfs\footnotemark\
\citep[]{obrienetal26-1,hollandsetal18-2} spanning a wide range of \Teff\ and
spectroscopic sub-types, and is of sufficient size to probe the statistical
properties of white dwarf chromaticities in general. To calculate chromaticities
for each of the white dwarfs in the 20\,pc volume, we used a mixture of
methodologies and sources of parameters depending on the spectral types, with
the results listed in Table~\ref{tab:twenty}.

\footnotetext{We do not include the purported cool, nearby white dwarf with
\emph{Gaia} DR3 ID 6013647666939138688 found by \citep{golovinetal24-1}, as its astrometry appears
contaminated, suggesting its large parallax may be erroneous.}

The majority of white dwarfs within 20\,pc have spectral types DA, DB, or DC.
For objects with these classifications we used pure hydrogen and pure helium
model grids spanning the same \Teff\ ranges as in Section~\ref{sec:DADBs}, but
instead of using a fixed $\log g$ of 8, we used extended grids spanning $\log g$
of 7.0 to 9.0 in steps of 0.5\,dex. In practice we found that their chromaticities
were almost entirely insensitive to $\log g$. For DCs where the atmospheric
composition may be ambiguous, we used the hydrogen/helium compositions of
\citet{obrienetal24-1}, which is our primary reference for 20\,pc white dwarf
parameters.

For the 20\,pc DQs we once again opted to use optical spectra rather than
models, which was particularly necessary for DQpec stars with distorted Swan
Bands. Since continuous spectral coverage over the entire visual range, and
accurate flux calibration is a necessity for converting spectroscopic data to
chromaticities, for most DQs we used spectral data hosted by the Montreal White
Dwarf Database \citep[MWDD,][]{dufouretal17-1} except where noted in
Table~\ref{tab:twenty}. We applied a similar adjustment to the flux calibrations
using PanSTARRS data to that described in Section~\ref{sec:DQs}.

For DZs within 20\,pc, we calculated spectral models using the
Koester white dwarf atmospheric model code, adopting atmospheric parameters
from the accompanying references in Table~\ref{tab:twenty}.
Note that for DAZs, the metal lines are sufficiently
narrow that they have a negligible effect on the calculated chromaticities,
and so we simply used DA models as above in those cases.

\begin{figure}
  \centering
  \includegraphics[width=\columnwidth]{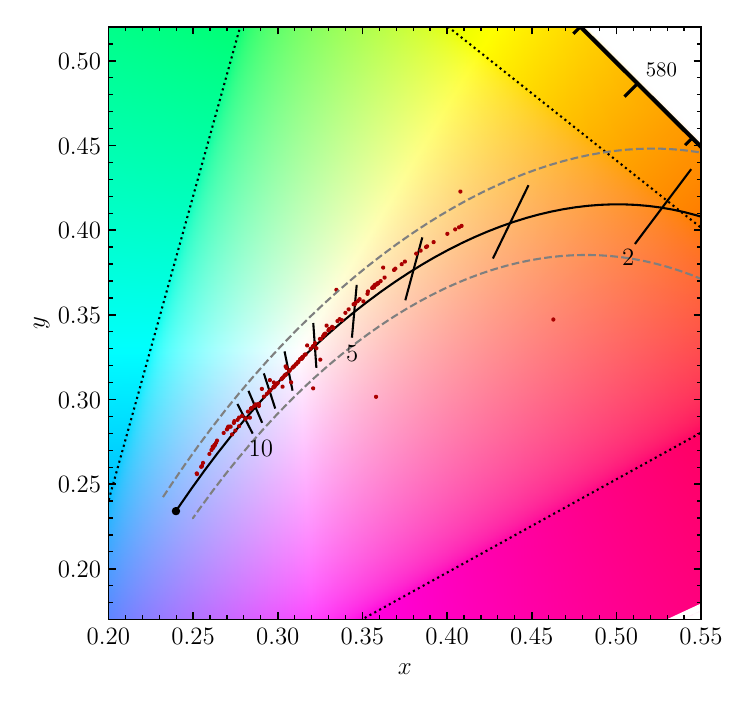}
  \includegraphics[width=\columnwidth]{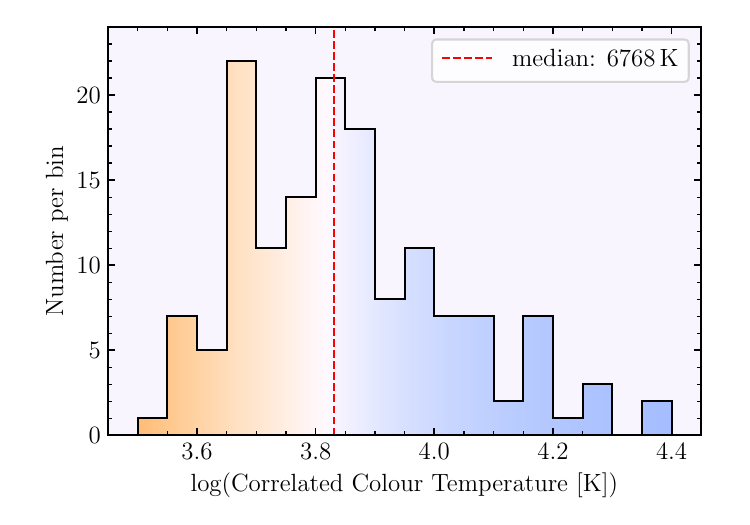}
  \caption{\label{fig:ctemp}
  Chromaticity distribution (top) for all white dwarfs within 20\,pc of the Sun
  (points). Our cuts to exclude objects with non-blackbody colours is shown by
  the grey dashed lines. The correlated colour temperature
  distribution for the 20\,pc sample (excluding three outliers) is shown in the
  bottom panel, where the colour gradient under the histogram represents the
  sRGB colour at each given colour temperature, and the panel background is
  filled with the colour corresponding to the median (\#F9F5FF).
    }
\end{figure}

The chromaticities for all white dwarfs within 20\,pc are shown in the top panel
of Fig.~\ref{fig:ctemp}. The majority of the 20\,pc white dwarfs have colours
close to the blackbody locus, including most of the DZs and DQs. Only a few
extreme stars deviate significantly away from blackbody colours. For objects
close to the blackbody locus it is simpler to examine the 1D distribution in
terms of correlated colour temperature. We calculated these
colour temperatures by converting the CIE 1931 $x$-$y$ chromaticities for each
object to UCS~1960 $u$-$v$ chromaticities using Equation~\eqref{eq:uv}, then
minimizing the Euclidean distance to the blackbody locus in the $u$-$v$ plane to
find the correlated colour temperature. For non-blackbody colours, e.g. green
and magenta, it is not meaningful to assign a colour temperature, and so we
excluded objects with distances greater than 0.01 (in $u$-$v$ terms) from the
Planckian-locus. The five rejected objects consisted of two DZs
(\wdj{0842}{-}{1347} and {\wdj{2253}{-}{0646}) both exhibiting yellow hues, and
three DQpec white dwarfs (\wdj{1011}{+}{2845}, \wdj{1038}{-}{2040}, and
\wdj{2157}{-}{5100}) all with pink colours. From this volume-limited sample, we
therefore conclude that only $3.4_{-1.3}^{+1.7}$\,percent of white dwarfs have
chromaticities deviating from blackbodies by more than the above threshold,
highlighting the rarity of the more unusual objects we identified in earlier
sections. 

The correlated colour temperature distribution of the remaining
147 objects are shown in the bottom panel of Fig.~\ref{fig:ctemp},
demonstrating a range in colour temperatures spanning 3000 to 25\,000\,K. The
median of this distribution, indicated by the vertical dashed line, is located
at 6768\,K with a corresponding sRGB hex code of \#F9F5FF.
Evidently this median colour is remarkably close to the D65 white point (a
colour temperature of 6504\,K), and so (to our amusement) the common depiction
of blue tinted white dwarfs is unrepresentative: on average white dwarfs are in
fact white, and are therefore appropriately named. Of course this neglects the
breadth of the distribution, where we find 95\,percent of objects are contained
within the range 3800 to 18\,700\,K, i.e. a factor of 5 in colour temperature.

\section{Visualisation of white dwarfs}
\label{sec:vis}

So far, all computed colours have been for observed spectra or disc-integrated
models only. The same techniques described in Section~\ref{sec:colourtheory} can
also be applied to spectral models to generate realistic images of stars as they
would appear to a nearby observer (though limited in brightness as before),
allowing the creation of images that can be used in an outreach setting.

The radiation emitted from the surface of a star is not only a function of
wavelength, $\lambda$, but also depends on the angle between the surface-normal
and the viewing direction, $\theta$. Therefore the fluxes $F$ can be expressed
as $F(\lambda, \mu)$ (the spectral radiance) where $\mu = \cos(\theta)$,
resulting in limb-darkening of stars, and is an important consideration in the
analysis of eclipsing binaries and exoplanet transits
\citep[e.g.][]{southworth12-1}. Limb-darkening must also be considered to generate
accurate images of stars, though not only its effect on brightness across the
stellar disc, but also its effect on chromaticity.

Using model atmospheres, the relative flux in various pass-bands can be
determined for different values of $\mu$. This is achieved by integrating
$F(\lambda, \mu)$ over a passband transmission curve, resulting in a function of
$\mu$ only, $I(\mu)$. These $I(\mu)$ are then typically fitted with a variety of
limb-darkening laws, with the fitted parameters tabulated for all relevant bands
and atmospheric parameters \citep[e.g.][]{gianninasetal13-1,claretetal20-2}. The
simplest limb-darkening law, the linear law,
\begin{equation}
    \frac{I(\mu)}{I(1)} = 1 - a(1-\mu),
\end{equation}
contains a single free parameter, $a$. However, in most cases, $I(\mu)$ departs
from a linear dependence on $\mu$ and so more complex limb-darkening laws are
required for accurate modelling. For instance, the quadratic limb-darkening law
extends the linear law by including a $(1-\mu)^2$ term with one further free
parameter. The Claret four-term law \citep{claret00-1},
\begin{equation}
    \frac{I(\mu)}{I(1)} = 1 - \sum_{k=1}^{4} a_k \left(1-\mu^{k/2}\right),
\end{equation}
provides a flexible function for modelling limb-darkening across a wide range of
\Teff\ and spectral types, and is the limb-darkening law we use here.

The Koester white dwarf model atmospheres internally calculate the spectral
radiance, $F(\lambda, \mu)$, at nine discrete values of $\mu$, which can be
output from the spectral synthesis allowing limb-darkening laws to be applied.
For a given spectral model, we calculated the XYZ values for each value of $\mu$
by integrating over $F(\lambda, \mu)$ using equation~\eqref{eq:ciexyz}. As
before we then converted these to linear sRGB using
equation~\eqref{eq:rgb_from_xyz}, with each channel now a function of $\mu$. We
then fitted limb-darkening laws to each of the R, G, and B channels, by
minimizing the squared differences between the limb-darkening model and the RGB
channel values. In our testing, only the Claret four-term law could reliably fit
all RGB curves without leaving excessive residuals. Finally we rescaled the
best-fitting limb-darkening curves so that brightest RGB value had an intensity
of 1 at $\mu=1$, maintaining the relative intensities between the $R$, $G$, and
$B$, thereby allowing linear sRGB values to be evaluated at any point on the stellar
disc.

We used our fitted and scaled limb darkening curves to generate images within
the unit circle, interpolating the fitted sRGB limb-darkening curves for each
pixel in the images. These images were then converted to display sRGB by
applying gamma-correction using equation~\eqref{eq:gamma_srgb}. Note that the
inclusion of gamma-correction is especially important here, as its omission
leads not only to incorrect chromaticities but also greatly exaggerated
limb-darkening.

We generated images for a small sample of DAs, DQs, and DZs which are shown in
the right panels of Figs.\,\ref{fig:visDA}--\ref{fig:visDZ2}. These are
accompanied by a selection of panels showing the spectra (top left), with the
bottom left panels demonstrating the fitted limb-darkening curves in terms of
sRGB intensity vs. $\mu$, sRGB intensity vs. $r$ (where $r^2 = 1-\mu^2$), and
chromaticity vs. $r$, from left to right.

For the DAs, we generated three models at $\Teff=7000$\,K, $10\,000$\,K and
$20\,000$\,K, all at $\log g=8$. While the linear intensities decrease by up to
80\,percent (in the 7000\,K model) from the centre to the limb, the change in
perceived intensity in the final images is fairly minor, due to the non-linear
response of the human visual system (Section~\ref{sec:gamma}), and because most
of the change occurs rapidly in the outermost pixels only. As a result, the
three DA images appear as almost flat discs. The chromaticity as a function of $r$
panels demonstrate that this quantity is almost constant across the entire
stellar disc with small changes seen at the very limbs. This results in a
slightly less intense shade of blue at the very edges of the stellar discs,
where the RGB intensities are close to equal near $\mu=0$ for the $7000$\,K and
$10\,000$\,K models.

In Sections\,\ref{sec:DQs} and \ref{sec:20pc} we restricted ourselves to
observed DQ spectra, in order to determine accurate chromaticities. However, to
generate images of these carbon-rich stars, it is necessary to use
angular-dependent model spectra. Using the Koester white dwarf model atmosphere
code, we have generated models for three DQs with increasingly complex Swan band
features, \wdj{0136}{+}{2004}, \wdj{0902}{+}{2010}, and \wdj{1011}{+}{2845}. The
greatest challenge in modelling these stars is capturing the distortion of the
C$_2$ Swan bands. \citet{kowalski10-1} demonstrated that this occurs due to
shifts in their electronic transition energy, $\Delta T_e$, that is proportional
to the atmospheric density, $\rho$, i.e. $\Delta T_e = \alpha \rho$, where
$\Delta T_e$ is in eV, and $\rho$ is in g\,cm$^{-3}$. Using density functional
theory methods \citet{kowalski10-1} found a value of $\alpha=1.6$, though this
severely overestimates the observed level of distortion. Instead
\citet{blouinetal19-3} determined an empirical value of $\alpha=0.2$, which was
found to produce acceptable results for many peculiar DQs, though still failed
in some extreme cases. Studying the evolution of DQ stars \citet{blouinetal19-4}
found that carbon-rich white dwarfs transition from DQs to DQpec when their
photospheric densities reach 0.15\,g\,cm$^{-3}$. We adopted the $\alpha=0.2$
calibrated by \citet{blouinetal19-3} in our models. For \wdj{0136}{+}{2004} we
adopted parameters from \citet{coutuetal19-1}, for \wdj{0902}{+}{2010} we used
parameters found by \citet{blouinetal19-4}, and for \wdj{1011}{+}{2845} we used
\citet{blouinetal19-3} parameters.

Our DQ images are shown in Fig.\,\ref{fig:visDQ}. In all three cases, limb
darkening is now readily apparent compared to the DAs, due to their atmospheres
being primarily composed of dense helium. \wdj{0136}{+}{4513} appears a pale
lilac colour in the centre of the disc, with a chromaticity shifting towards
fainter and more pink hues at the limb of the star. The image for the DQpec
\wdj{0902}{+}{2010} is pink throughout, though the colour becomes less vivid
towards the limb. Despite some minor differences between the model and SDSS
spectrum (the ridges between the Swan bands are slightly more pronounced in the
model), the overall colour is fairly similar to that found in
Fig.~\ref{fig:DQspec1}. Finally we show the reddish-pink DQpec
\wdj{1011}{+}{2845}. Like \citet{blouinetal19-3}, our model does not fully
reproduce the roundness of the Swan bands, which are much more distorted in the
data (Fig.~\ref{fig:DQspec1}), possibly due to the presence of a magnetic
field \citep{schmidtetal99-1}. Overall this results in an disc-averaged model
colour (\#FF7A65) which is more vivid than the observed spectrum (\#FF8775),
though the model still retains the salmon pink hue found from the data.
Like the other DQs the limb-darkening is quite pronounced, and also shows a
shift to a less vivid tone towards the disc edge. 

For the DZs we generated images for six objects in order demonstrate the wider
range of chromaticities for these stars, which are shown in
Figs.~\ref{fig:visDZ1} and \ref{fig:visDZ2}. The first of these
\wdj{1535}{+}{1247} has a typical cool DZ spectrum, and is a well studied
\citep{koesteretal11-1,hollandsetal17-1,blouinetal19-3,coutuetal19-1,hollandsetal22-1}
member of the 20\,pc sample \citep{hollandsetal18-2}. Owing to its near-Solar
\Teff\ the centre of the disc is close in colour to the D65 white point, i.e.
all three RGB values are close to 1 for $\mu=1$. However, limb-darkening induces
a golden hue around the edge of the star. \wdj{0916}{+}{2540}, first analysed
by \citet{koesteretal11-1} has a complex spectrum resulting from large metal
abundances, giving it a pale yellow-orange colour. While limb-darkening in its
image is apparent, the chromaticity shift is more subtle for this object.
\wdj{0744}{+}{4649} is a similar case, with its vivid yellow colour remaining near
constant in chromaticity across the disc, where the $R$ and $G$ values are near
equal over all values of $\mu$.

We show two green objects in Fig.~\ref{fig:visDZ2}. The first of these,
\wdj{0851}{+}{1543} is relatively pale, but shows a complex change in
chromaticity across the disc. At the disc centre, the $G$ and $B$ values are
virtually identical (pale cyan), but becomes $G$ dominated for $\mu < 0.9$
($r>0.4$). However, for $\mu < 0.15$ ($r>0.99$), $R$ becomes the dominant
channel, giving a slight yellow hue to the very limb of \wdj{0851}{+}{1543}.
\wdj{1330}{+}{6435} exhibits a more vivid green colour (\#D0FFB4) at the disc
centre, though the colour becomes less saturated towards the limbs before
becoming $R$ dominated for $\mu <0.2$ ($r>0.98$). Upon close inspection of the
image, this gives \wdj{1330}{+}{6435} an orange hue in the outermost pixels of
its disc. Finally, we show \wdj{2356}{-}{2054}, which (as discussed in
Section~\ref{sec:DZs}) is essentially a more metal-rich analogue of
\wdj{1330}{+}{6435} with an even larger sodium abundance. Despite being only
4000\,K, the star is a vivid blue across its entire disc. Unlike the DA models
shown in Fig.~\ref{fig:visDA}, which are blue due to their high \Teff,
\wdj{2356}{-}{2054} shows strong limb darkening, betraying the nature of its
cool helium-dominated atmosphere. 

Images such as those shown in Figs.\,\ref{fig:visDA}--\ref{fig:visDZ2} show
that the appearances of white dwarfs are highly diverse. This is shown to be
particularly true for cool white dwarfs with helium dominated-atmospheres
enriched in heavy elements where the strong opacity sources results in not only
unusual colours, but also chromaticity shifts across the stellar disc due to limb
darkening.

\section{Conclusions}
\label{sec:conc}

We have shown that white dwarfs exhibit a wide range of chromaticities. We found
that the chemically simplest white dwarfs with pure hydrogen or helium
atmospheres closely follow the blackbody locus in the chromaticity diagram. In
volume-limited samples, we established that white dwarfs have a median
correlated colour temperature of 6800\,K, and so on average white dwarfs are
indeed white in colour, rather than the canonical blue hue corresponding to hot
white dwarfs which are over-represented in magnitude-limited samples.

More exotic white dwarfs, i.e. those with carbon or other metals in their
atmospheres, may exhibit extremely broad absorption features at optical
wavelengths. For DQs we found this results in pink and purple stars, whereas for
DZs we identified examples known objects with yellow and green hues.
Furthermore, IR faint white dwarfs with mixed hydrogen/helium atmospheres turn
towards cyan colours due to collision induced absorption of H$_2$-He encroaching
into optical wavelengths.

Finally, we demonstrated that calculations typically used to determine
limb-darkening coefficients from model atmospheres can be applied to generate
realistic images of stars. In the case of DQ and DZ white dwarfs, the dramatic
limb-darkening can lead to strong variations in both perceived intensity and
chromaticity across the stellar disc. We expect such images to be useful in an
outreach setting, highlighting the non-blackbody radiation emitted from stellar
atmospheres under extreme physical conditions.

\section*{Acknowledgements}

The authors thank the referee, Patrick Dufour, for their constructive
feedback. MAH acknowledges a Warwick Astrophysics prize post-doctoral
fellowship made possible thanks to a generous philanthropic donation. PET and
BTG received funding from the European Research Council under the European
Union’s Horizon 2020 research and innovation programme number 101002408
(MOS100PC) and 101020057 (WDPLANETS). DK acknowledges V. Alekseev for Na-He
adiabatic potentials and dipole moments used for Na unified profiles. MAH
acknowledges M.~W. O'Brien for the use of DZ models from the 13\,pc sample, A.
Stockman for useful conversations about colour matching functions, and J.~J.
Hermes, R.~P. Ashley, and C.~E. Pugh for continued encouragement to publish this
unconventional analysis.

\section*{Data Availability}

All observed spectra were obtained from publicly available SDSS data or from the
Montreal White Dwarf Database. Spectral models for individual objects are
available upon reasonable request.

\section*{Conflicts of Interest}

The authors declare no conflict of interest.






\appendix

\section{Luminance}
\label{sec:lum}

Throughout this work, we have largely ignored the perceived brightness for all
objects we have considered, as computer displays are (thankfully) many orders of
magnitude fainter than the surfaces of stars. Therefore all displayed colours
and images are generated to be as bright as the sRGB standard allows, while
still preserving chromaticity. For completeness, we demonstrate here how the
perceived brightness can also be calculated from spectral models.

In section~\ref{sec:colourtheory}, we noted that $\CMF{y}$ is equivalent to the
photopic luminous efficiency function, i.e. the overall spectral sensitivity of
the human eye in photopic conditions. Therefore the integral defining $Y$ 
(equation\,\eqref{eq:ciexyz}) can be used to represent brightness. More
specifically, $Y$ can be defined as one of several luminous quantities according
to a modified version of equation~\eqref{eq:ciexyz} scaled by a prefactor\footnotemark\
\citep{candela},
\begin{equation}
    Y = 683.002\,\frac{\mathrm{lm}}{\mathrm{W}} \int_{360}^{830} F_\lambda(\lambda) \cdot\CMF{y}\,\mathrm{d}\lambda,
    \label{eq:lum}
\end{equation}
where $\lambda$ is again in nm and 1\,lm (short for 1\,lumen) is the unit of
luminous flux. The specific quantity and units of $Y$ therefore depend on the
units of $F_\lambda$.

\footnotetext{The luminous efficacy constant defining the candela,
$K_\mathrm{cd}$, is defined to be exactly 683\,lm\,W$^{-1}$ for a frequency of
540\,THz, with a corresponding air wavelength of 555.017\,nm. Because the CIE
photopic luminous efficiency function, $\CMF{y}$, instead peaks at 555\,nm
exactly, the scaling factor in equation \eqref{eq:lum} is slightly higher than
$K_\mathrm{cd}$.}

If $F_\lambda$ is the spectral irradiance of an observed star with units
W\,m$^{-2}$\,nm$^{-1}$ (or more typically erg\,s$^{-1}$\,cm$^{-2}$\,\AA$^{-1}$
in astronomy), then $Y$ is the illuminance, with units of lm\,m$^{-2}$ (or
equivalently lux). For a flat spectrum in $F_\lambda$ with a \emph{Gaia}
apparent magnitude of $G=10$ (similar to some of the brightest nearby white
dwarfs), the illuminance is a mere $Y=1.85\times10^{-10}$\,lux. If instead
$F_\lambda$ represents the spectral flux density radiated from the surface of a
star (with the same units), then $Y$ represents the luminous exitance, and is
again measured in lm\,m$^{-2}$ (lux is not used). However, if we wish to
describe how bright some point appears on the surface of resolved object
(independent of its size or distance to the observer), we must consider the
directional component of the radiated flux (as in Section~\ref{sec:vis}).
Therefore if $F_\lambda$ is the spectral radiance with units equivalent to
W\,m$^{-2}$\,sr$^{-1}$\,nm$^{-1}$, then $Y$ is termed the luminance, with units
lm\,sr$^{-1}$\,m$^{-2}$. More typically the luminance is expressed in equivalent
units\footnotemark\ of cd\,m$^{-2}$, where the SI base unit of luminous
intensity, the candela (cd), is equivalent to lm\,sr$^{-1}$.

\footnotetext{For display technology the ``nit'' is often used as an equivalent
unit for the luminance.}

The calculation of synthetic stellar spectra (such as the Koester white dwarf
models) requires the spectral radiance to be calculated at a variety of angles
in order to integrate over the stellar disc and compare with observations. We are
therefore able to calculate the luminance across the stellar disc for our models
shown in Section~\ref{sec:vis}. Using equation~\eqref{eq:lum}, we have
calculated the luminance at the centre of the stellar disc for a few of the
objects in Appendix Figs.~\ref{fig:visDA}--\ref{fig:visDZ2}. The brightest
object shown is the $20\,000$\,K DA model with a central luminance of
$5.1\times10^{10}$\,cd\,m$^{-2}$, whereas the faintest is \wdj{2356}{-}{2054} at
$2.3\times10^8$\,cd\,m$^{-2}$. For \wdj{1535}{+}{1247}, the central luminance is
$2.8\times10^{9}$\,cd\,m$^{-2}$ which is broadly similar to the luminance of the
Solar disc without attenuation from the Earth's atmosphere, owing to the similar
$\Teff$ of this white dwarf to the Sun. In comparison, standard dynamic range
computer displays (SDR) have peak luminances of a few 100\,cd\,m$^{-2}$,
whereas modern high dynamic range (HDR) displays can reach several
1000\,cd\,m$^{-2}$. Evidently, the brightest computer displays are vastly
outshone by even the faintest white dwarfs.

That being said, a luminance of 1000\,cd\,m$^{-2}$ corresponds to an
approximately 1330\,K blackbody. Therefore for a subset of brown dwarfs, images
could be generated with accurate luminances, without scaling to the limits of
the display. Of course the optical tails of brown dwarf spectra will appear
different to blackbodies \citep{bd_burrows01}, and so the actual \Teff-range
without luminance-scaling would require analysis of brown dwarf spectral models
which is beyond the scope of this work.

\section{Additional figures}
\label{sec:app_fig}
\begin{figure*}
  \centering
  \includegraphics[width=\textwidth]{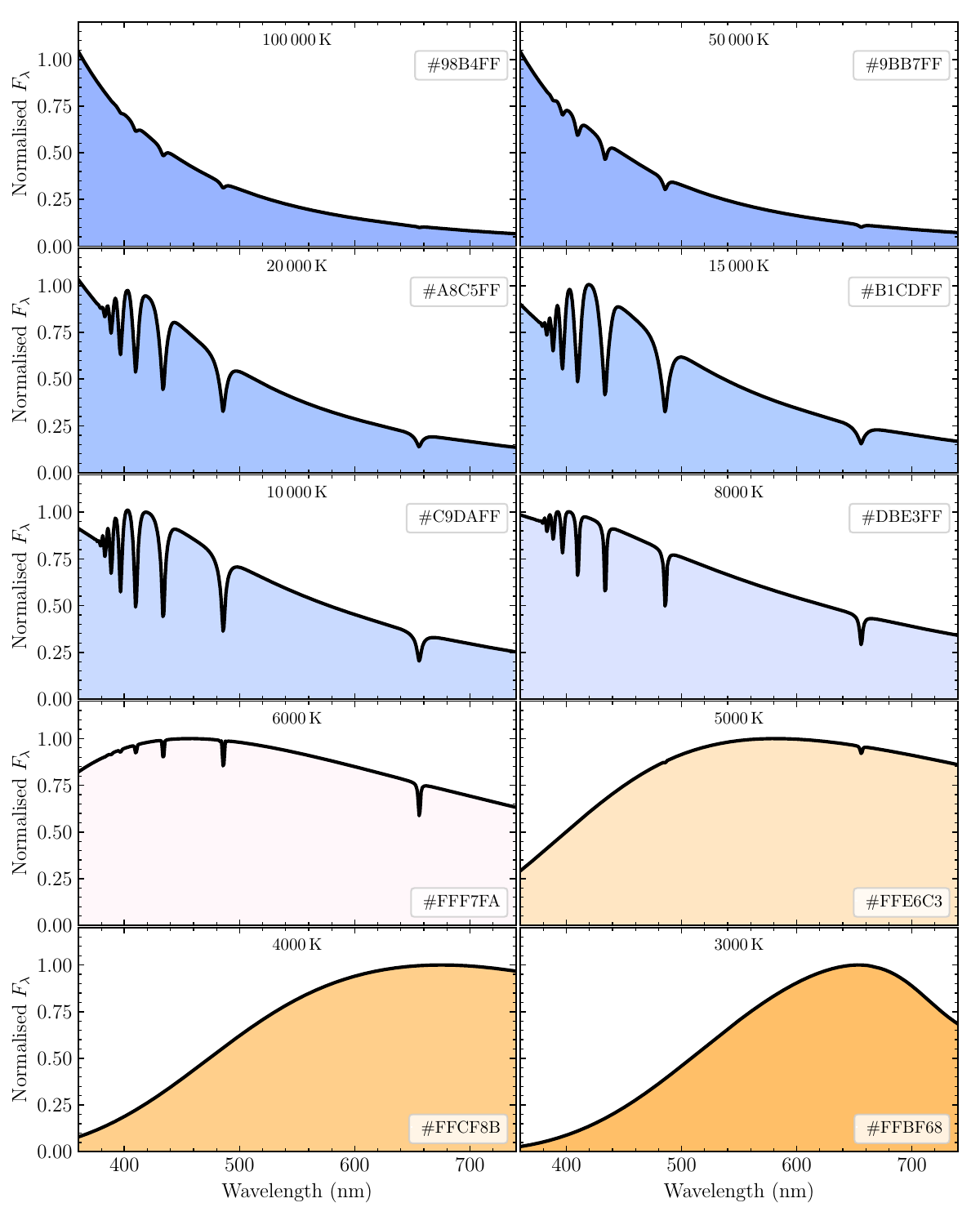}
  \caption{\label{fig:DAspec}
  A sample of pure hydrogen (DA) model spectra of different \Teff. Each spectrum is filled
  underneath with its calculated sRGB colour with the corresponding sRGB hexcode
  over-plotted on the figure.
    }
\end{figure*}

\begin{figure*}
  \centering
  \includegraphics[width=\textwidth]{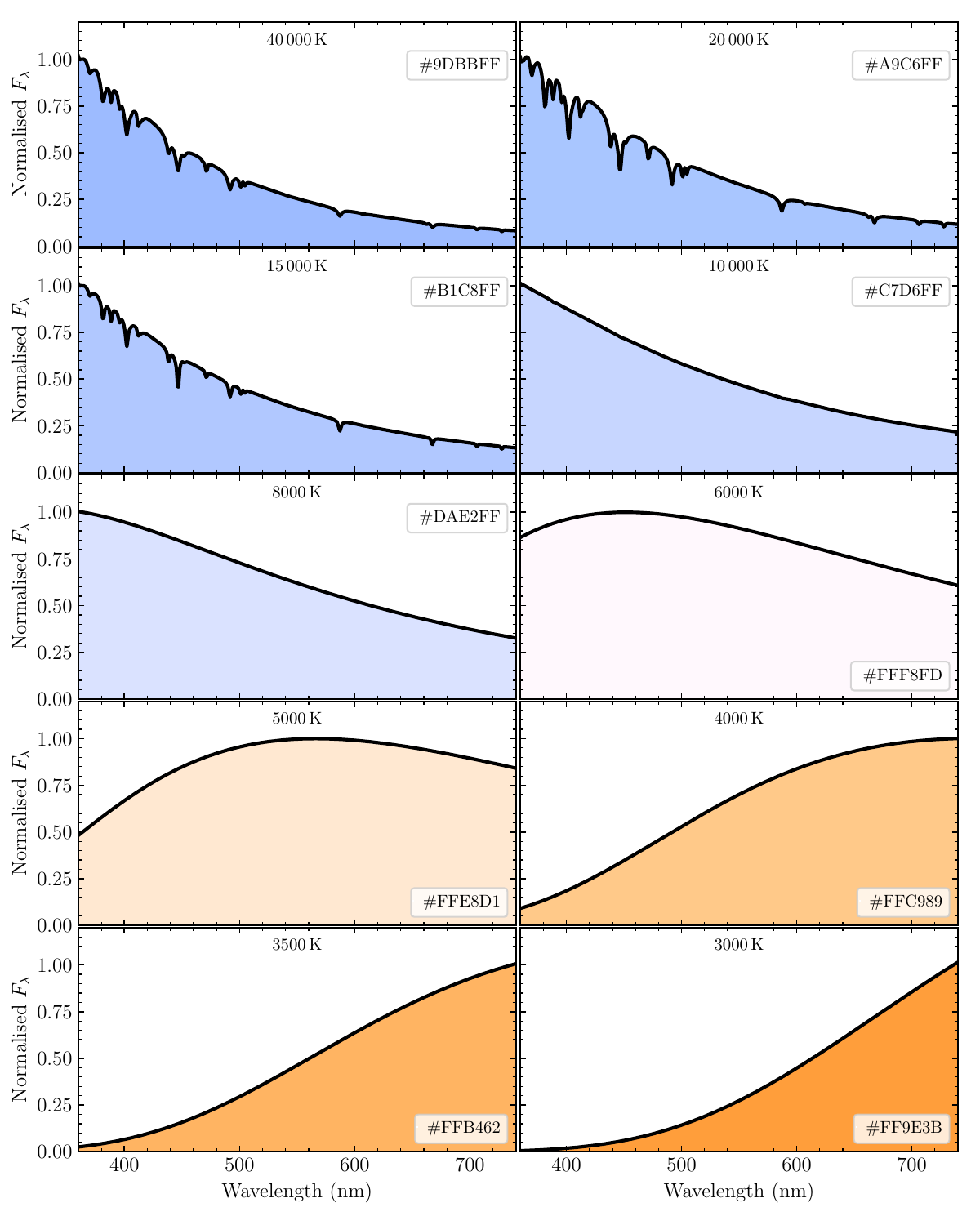}
  \caption{\label{fig:DBspec}
  Same as figure\,\ref{fig:DAspec}, but for pure helium (DBs) instead.
    }
\end{figure*}

\begin{figure*}
  \centering
  \includegraphics[width=\textwidth]{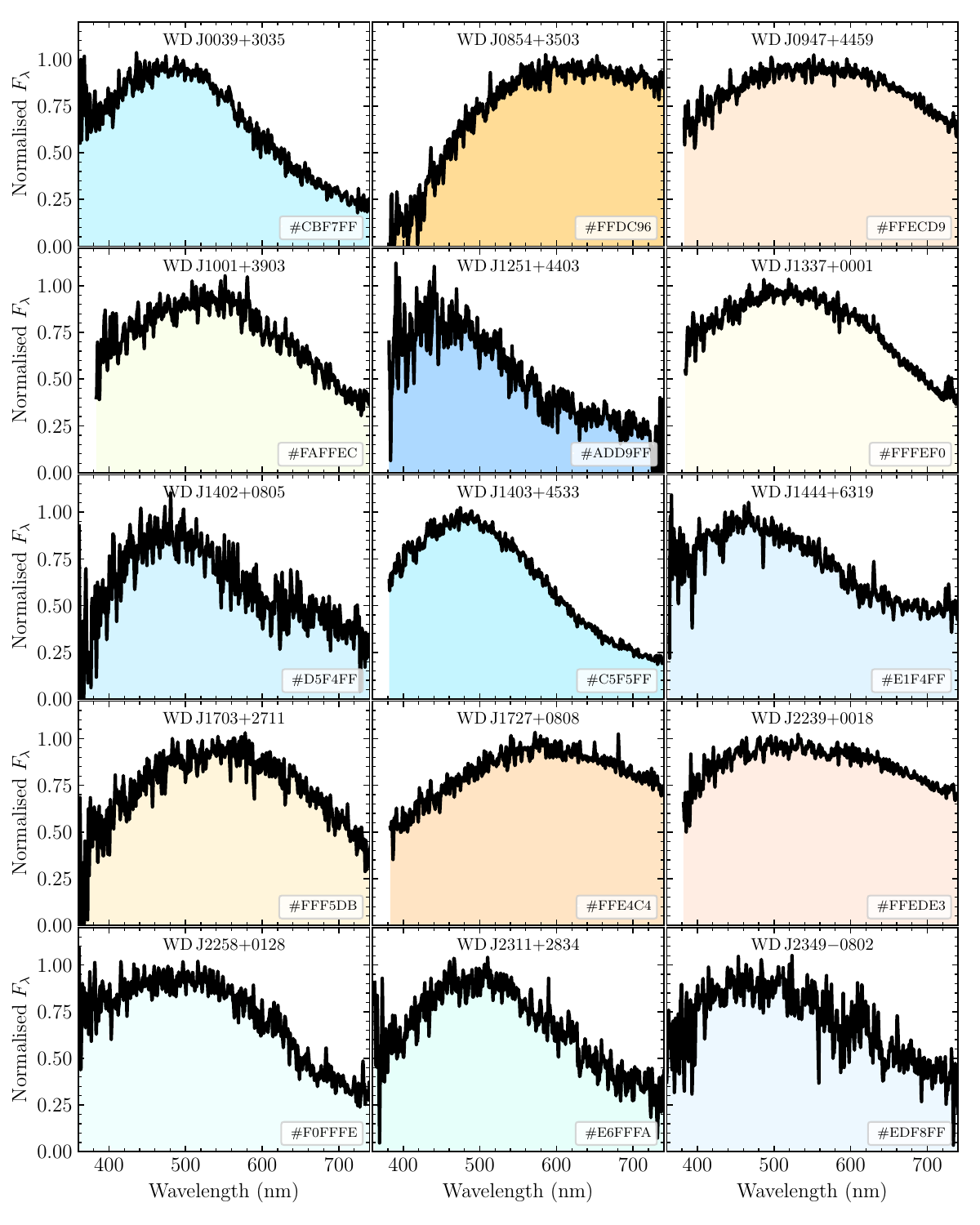}
  \caption{\label{fig:UCspec}
  Same as figure\,\ref{fig:DAspec}, but for IR-faint white dwarfs instead.
  Full coordinates are given in Table\,\ref{tab:UC}.
    }
\end{figure*}

\begin{figure*}
  \centering
  \includegraphics[width=\textwidth]{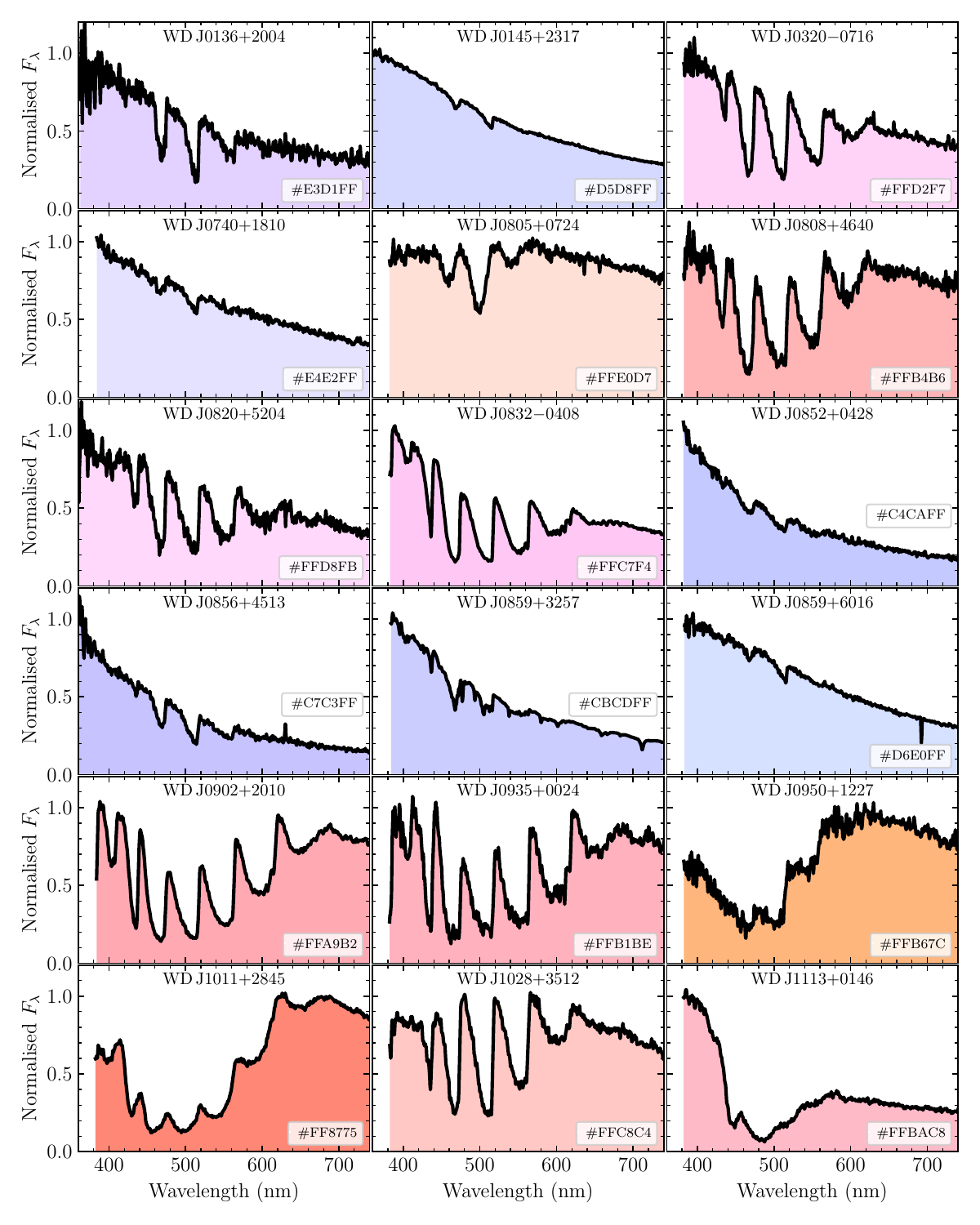}
  \caption{\label{fig:DQspec1}
  Same as figure\,\ref{fig:DAspec}, but for DQs instead.
  Full coordinates are given in Table\,\ref{tab:DQ}.
    }
\end{figure*}

\begin{figure*}
  \centering
  \includegraphics[width=\textwidth]{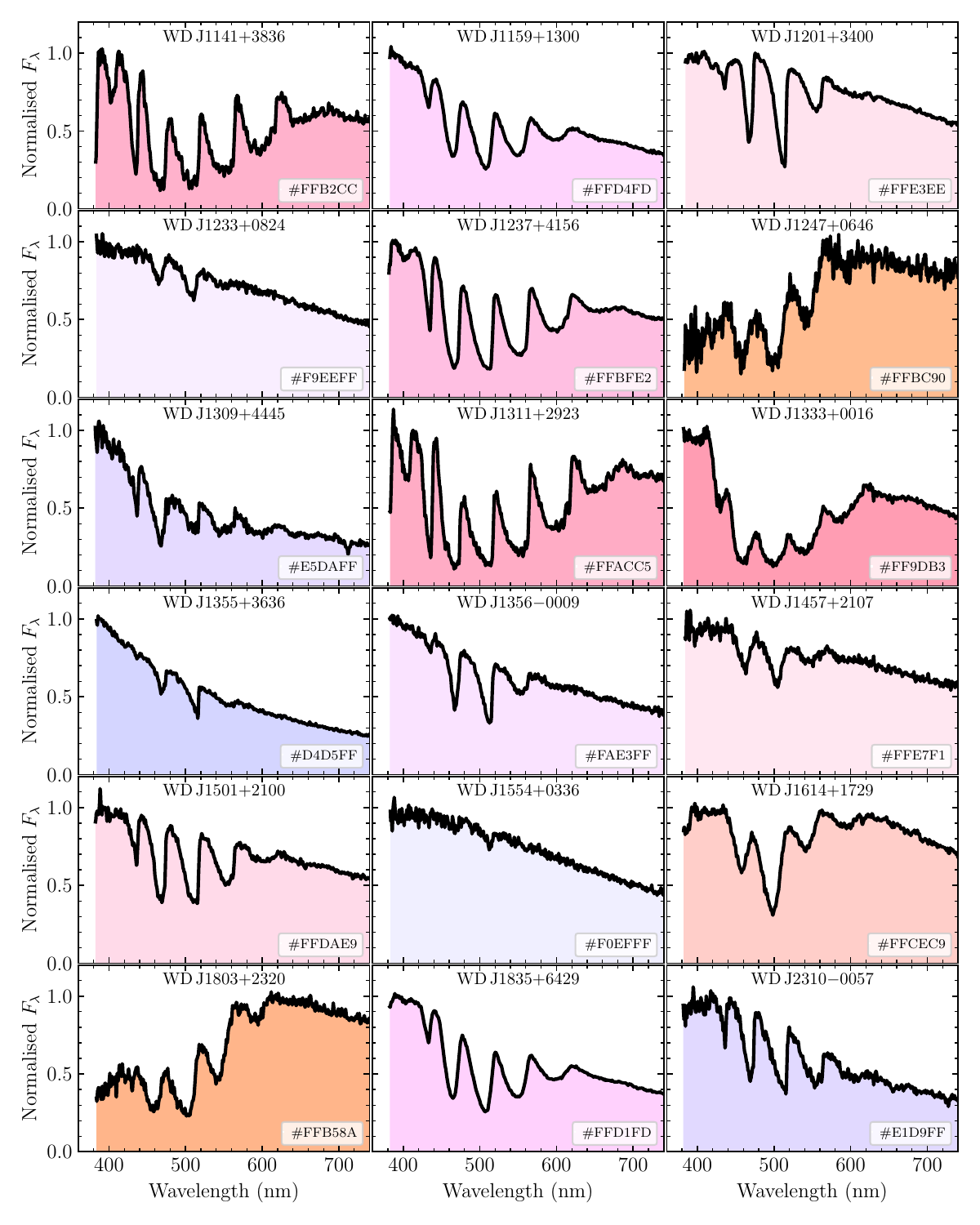}
  \caption{\label{fig:DQspec2}
  Additional DQ spectra.
    }
\end{figure*}

\begin{figure*}
  \centering
  \includegraphics[width=\textwidth]{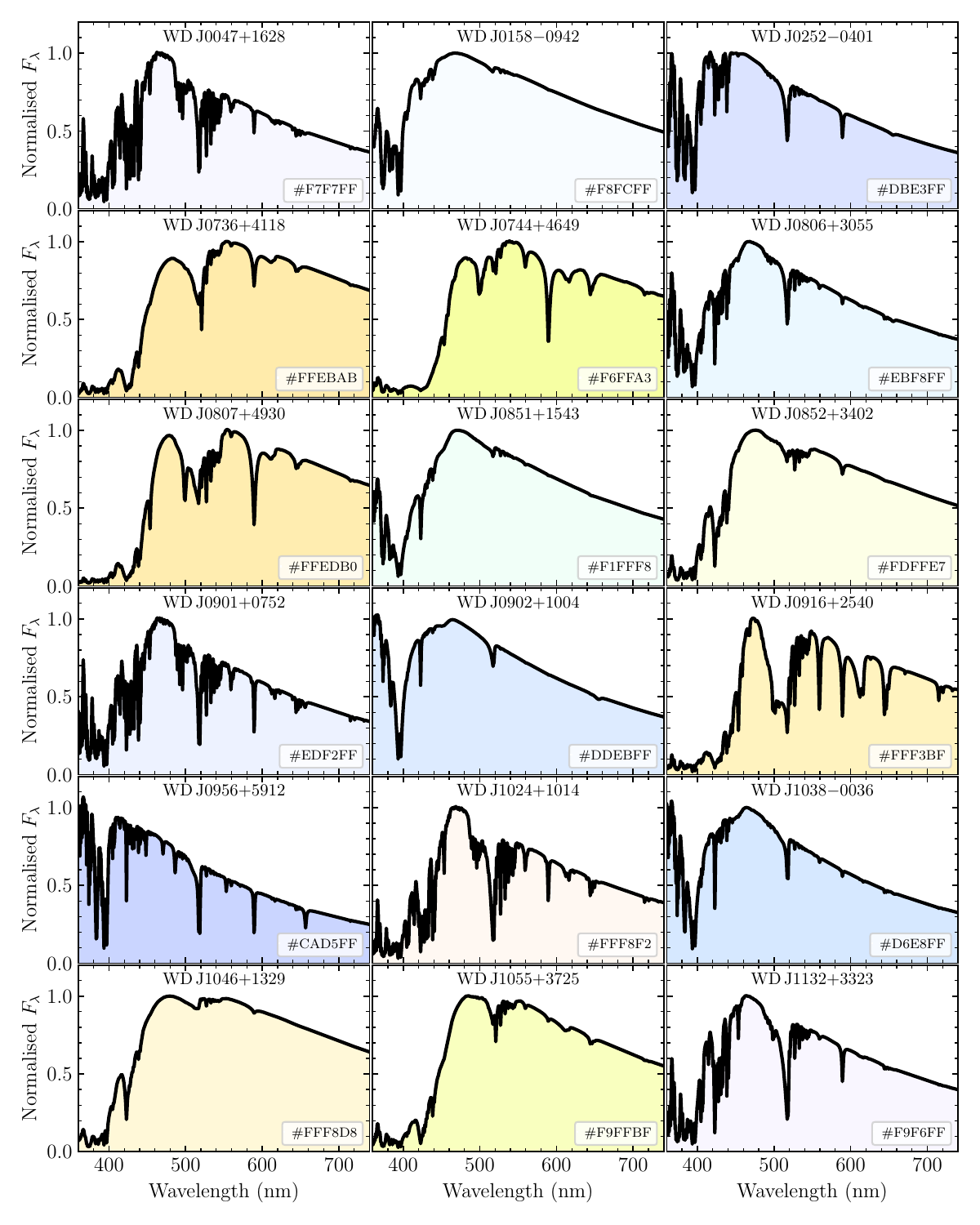}
  \caption{\label{fig:DZspec1}
  Same as figure\,\ref{fig:DAspec}, but for DZs instead.
  Full coordinates are given in Table\,\ref{tab:DZ}.
    }
\end{figure*}

\begin{figure*}
  \centering
  \includegraphics[width=\textwidth]{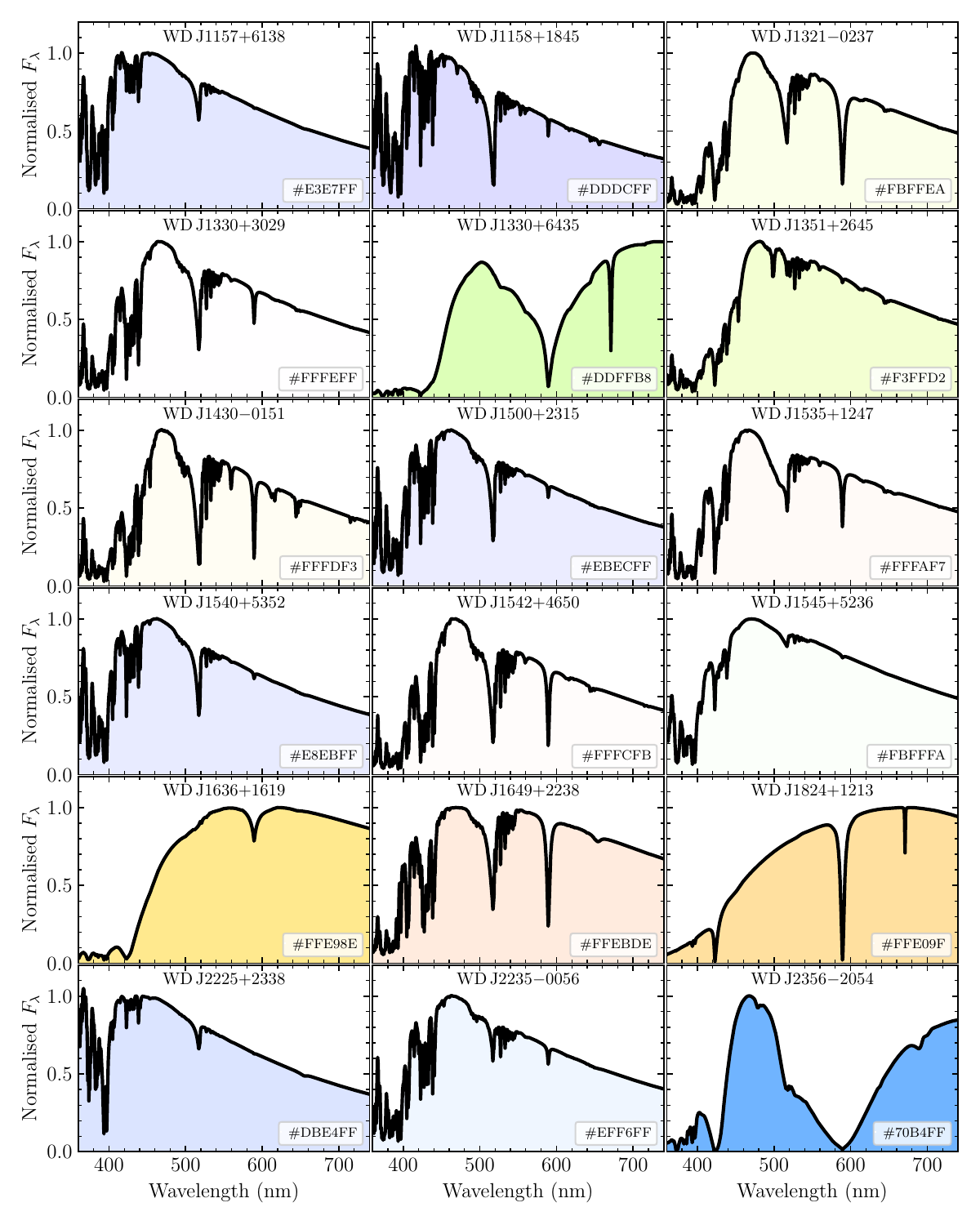}
  \caption{\label{fig:DZspec2}
  Additional DZ spectra.
    }
\end{figure*}

\vspace{2cm}

\begin{figure*}
  \centering
  \includegraphics[width=\textwidth]{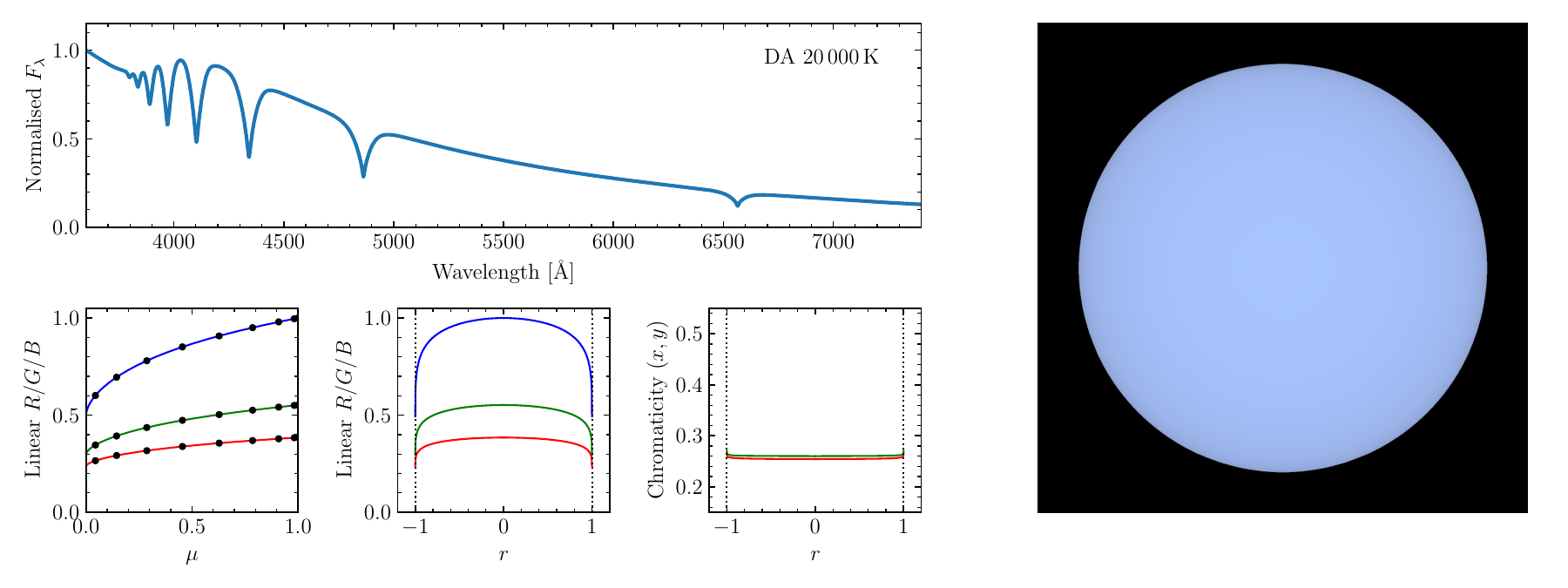}
  \includegraphics[width=\textwidth]{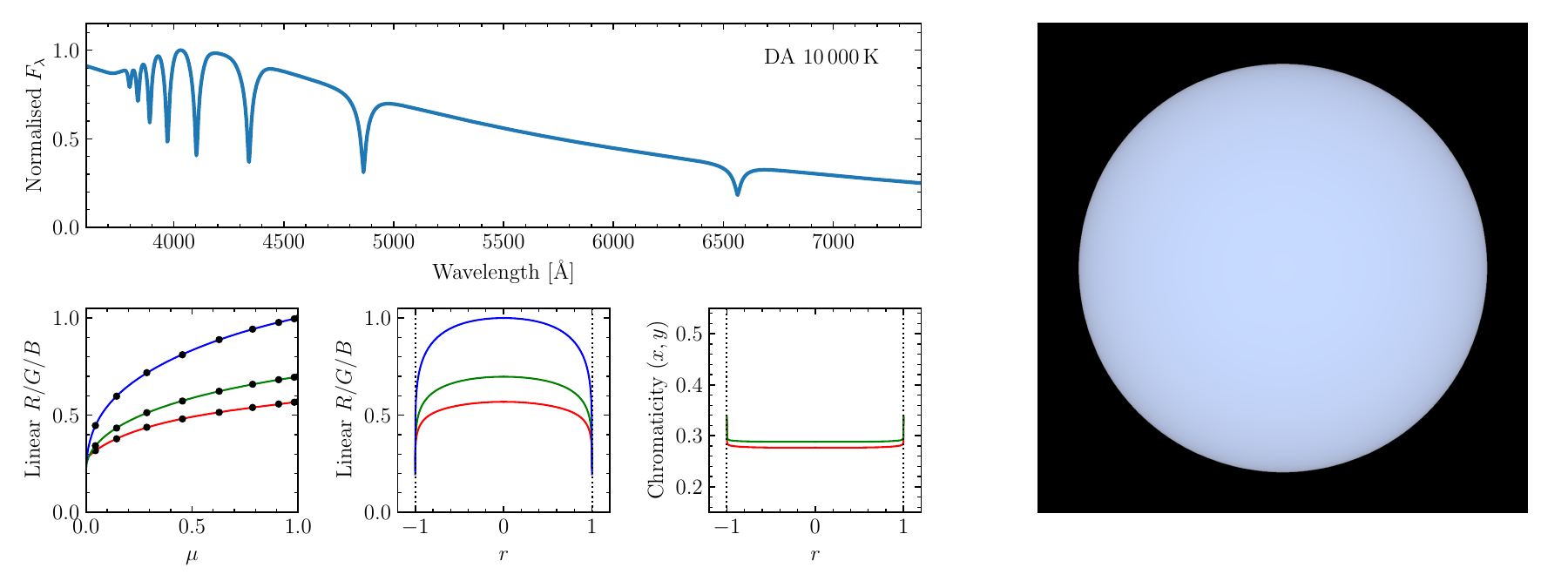}
  \includegraphics[width=\textwidth]{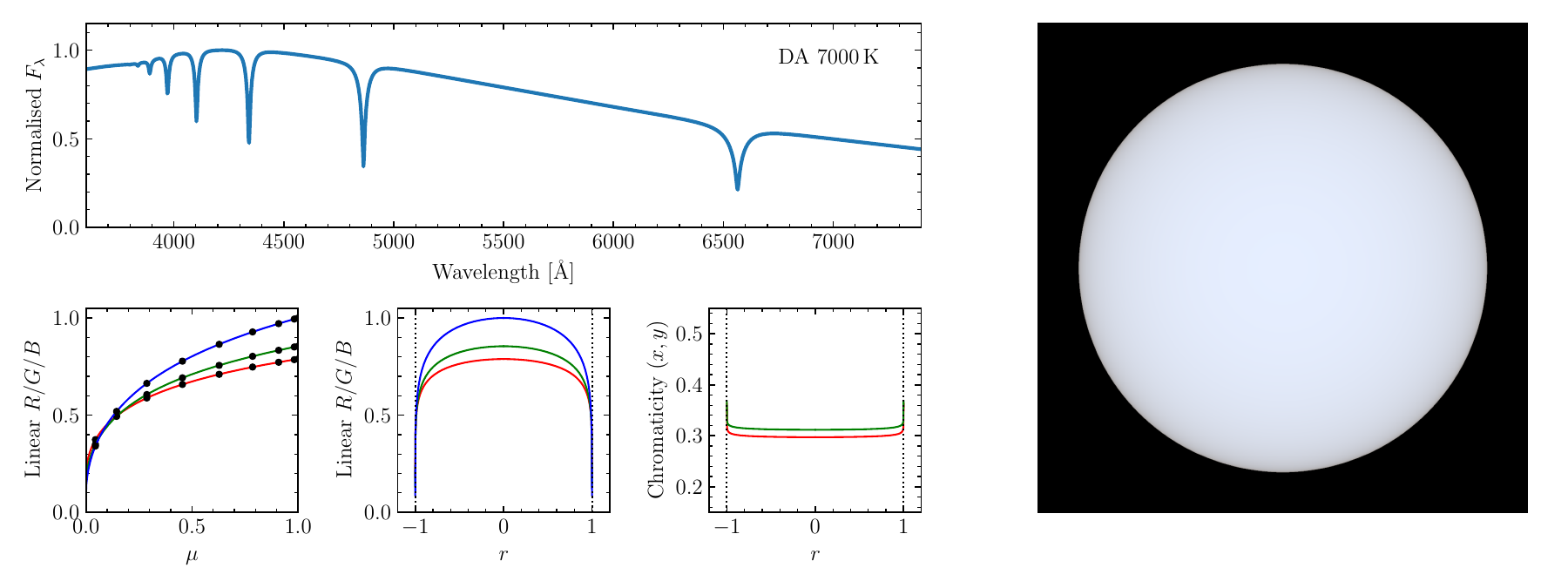}
  \caption{\label{fig:visDA}
    Visualisations for three DA white dwarf models at different \Teff\ (right). The bottom left panels (per object)
    show the relative intensities in the sRGB channels (red, green, and blue curves respectively) as a function
    of $\mu$, with Claret 4-parameter curves fitted to the values from the model atmospheres (black dots).
    The bottom middle panels show the same the same linear intensities as a function of the projected radius, $r$.
    The bottom right panels instead show the chromaticity as a function of $r$, with red and green curves corresponding
    to $x$ and $y$ values respectively. 
    The top left panels show the disc-averaged spectra used to generate the images.
    }
\end{figure*}

\begin{figure*}
  \centering
  \includegraphics[width=\textwidth]{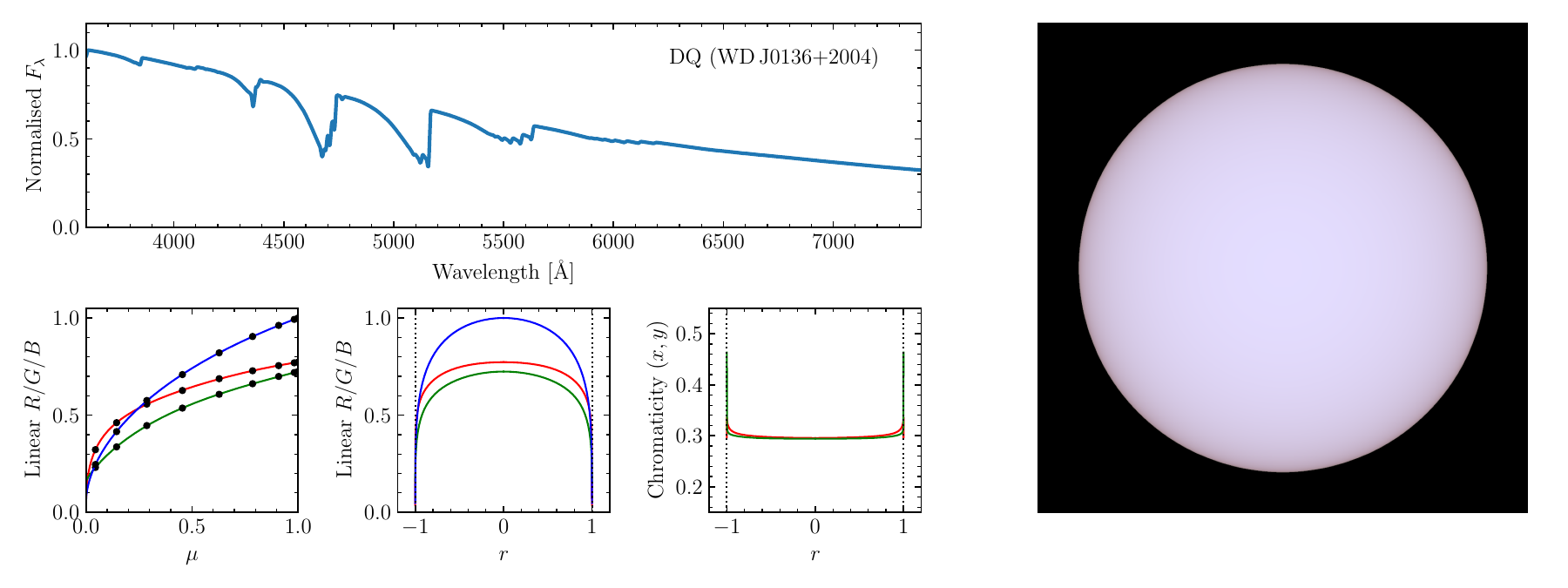}
  \includegraphics[width=\textwidth]{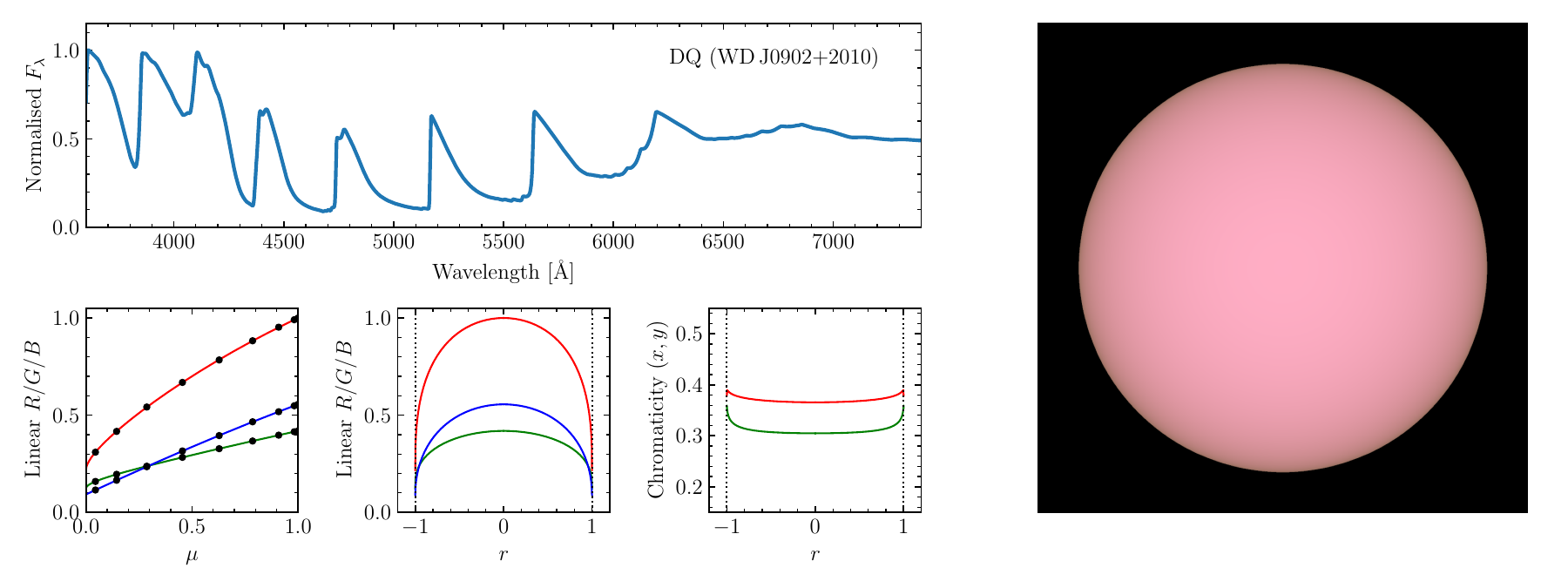}
  \includegraphics[width=\textwidth]{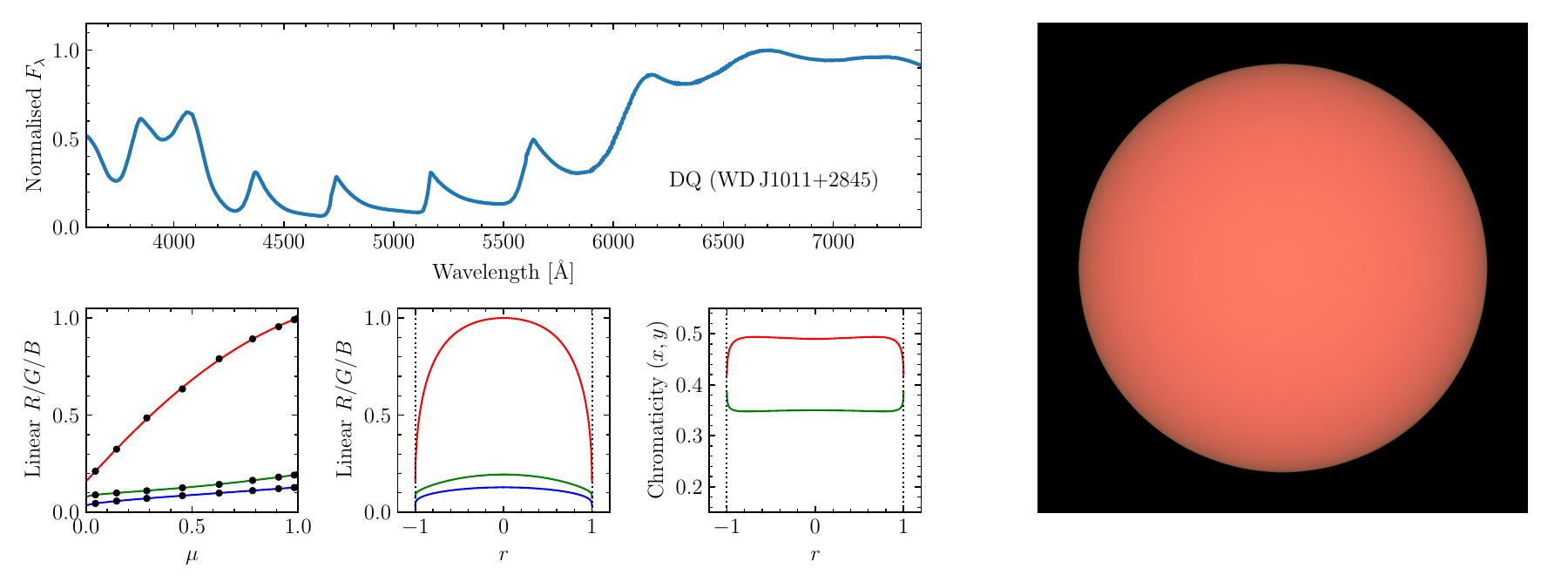}
  \caption{\label{fig:visDQ}
    Visualisations for three DQ white dwarf models. Panels to the left of the main images have the same meaning
    as in Figure\,\ref{fig:visDA}.
    }
\end{figure*}

\begin{figure*}
  \centering
  \includegraphics[width=\textwidth]{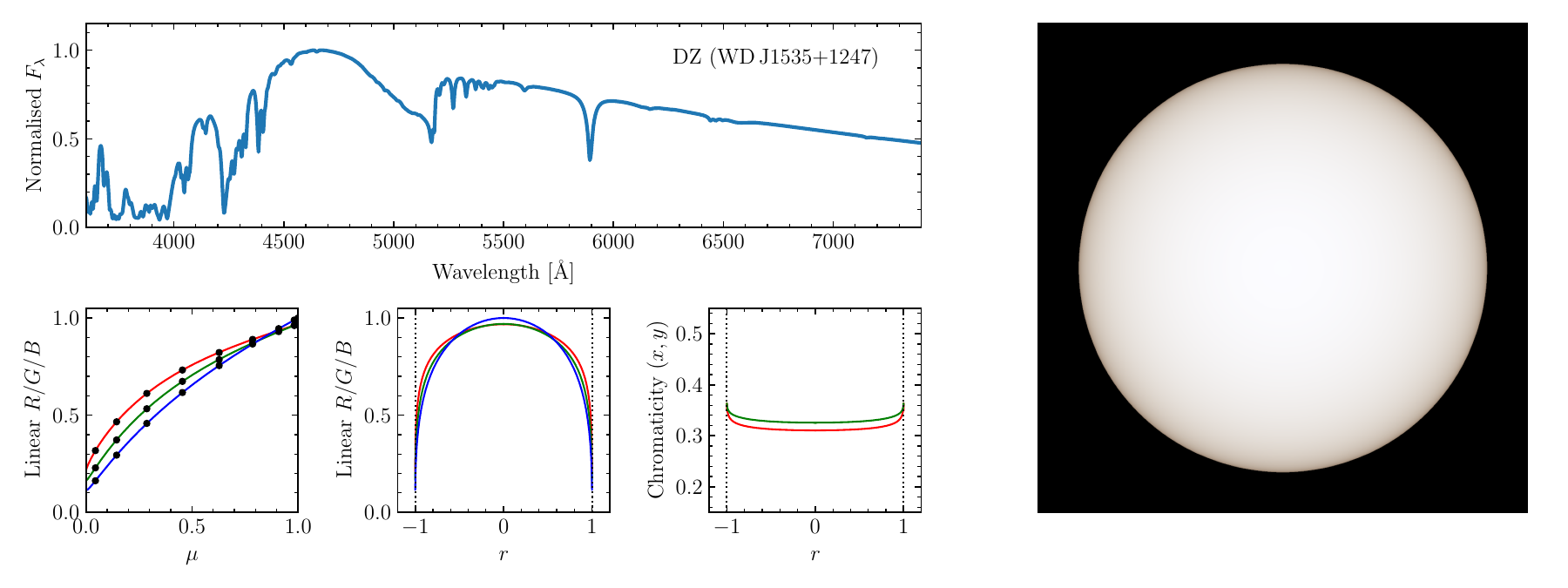}
  \includegraphics[width=\textwidth]{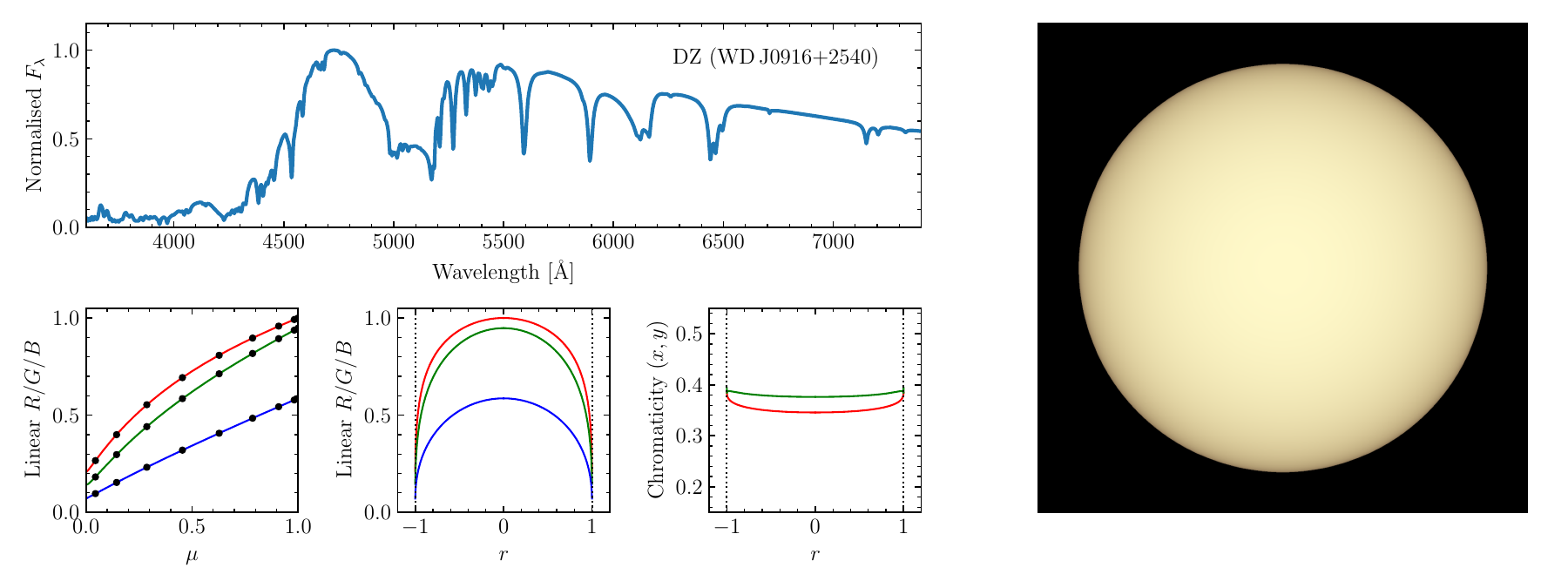}
  \includegraphics[width=\textwidth]{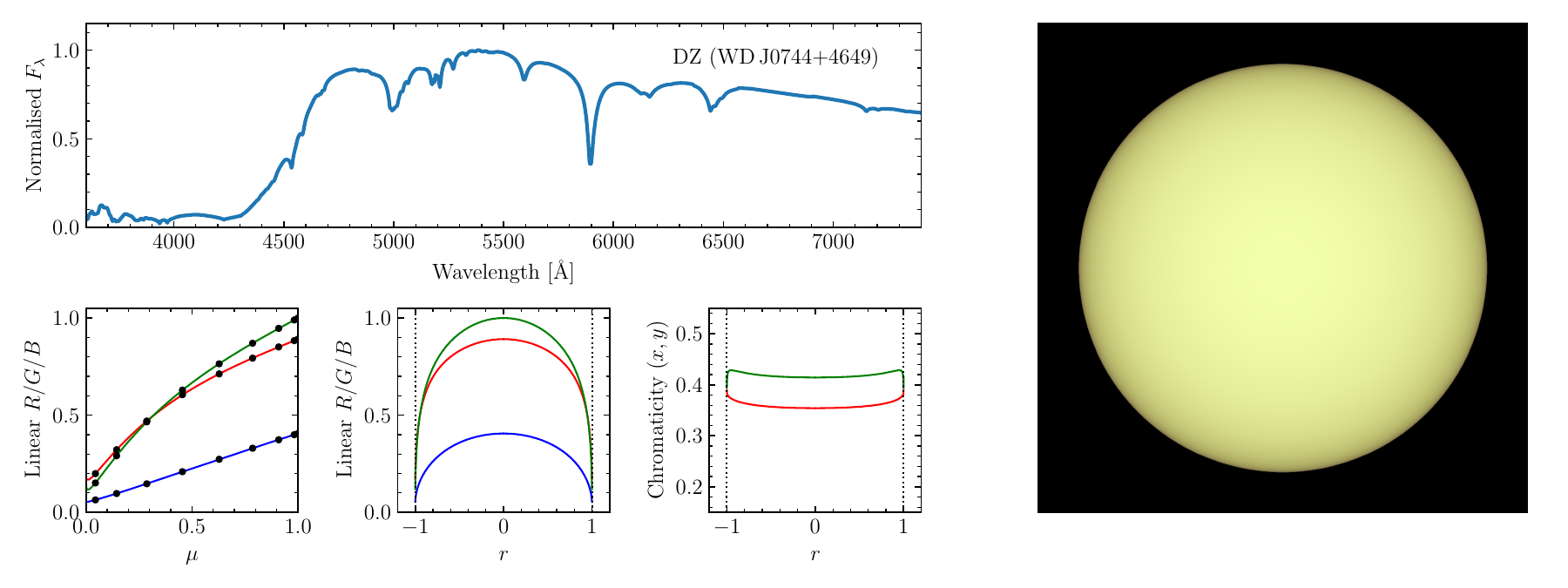}
  \caption{\label{fig:visDZ1}
    Visualisations for three DZ white dwarf models. Panels to the left of the main images have the same meaning
    as in Figure\,\ref{fig:visDA}.
    }
\end{figure*}

\begin{figure*}
  \centering
  \includegraphics[width=\textwidth]{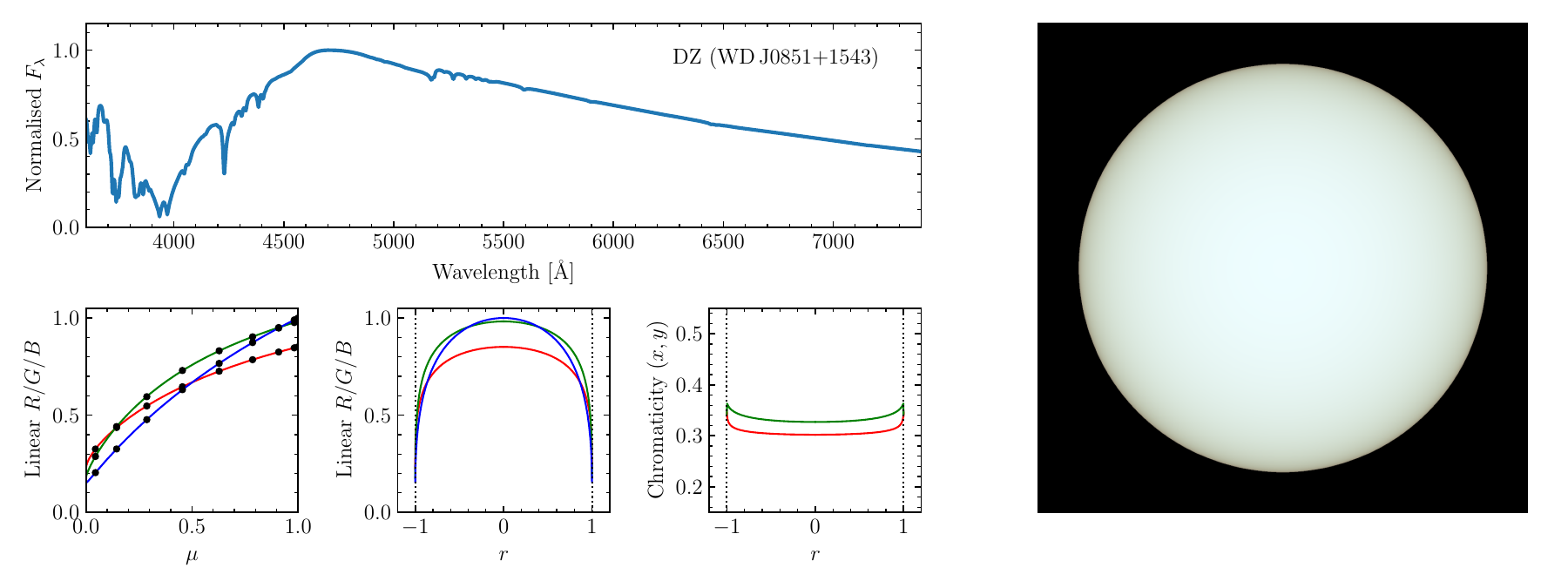}
  \includegraphics[width=\textwidth]{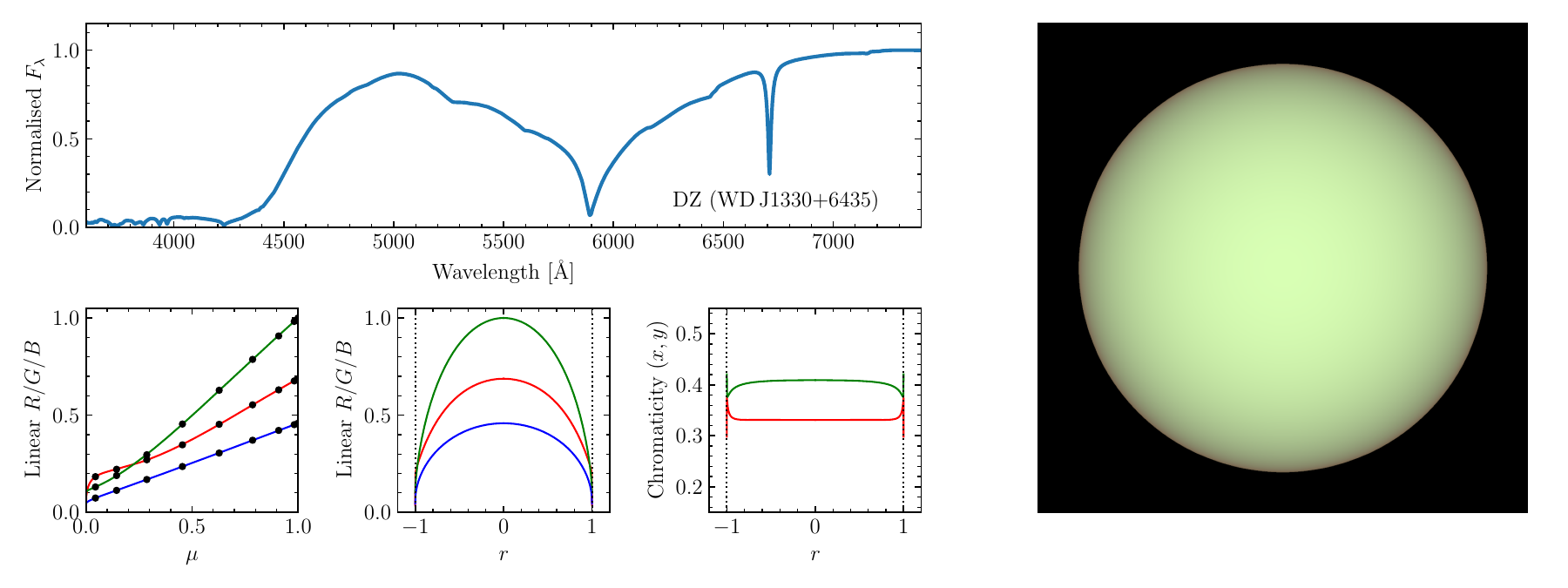}
  \includegraphics[width=\textwidth]{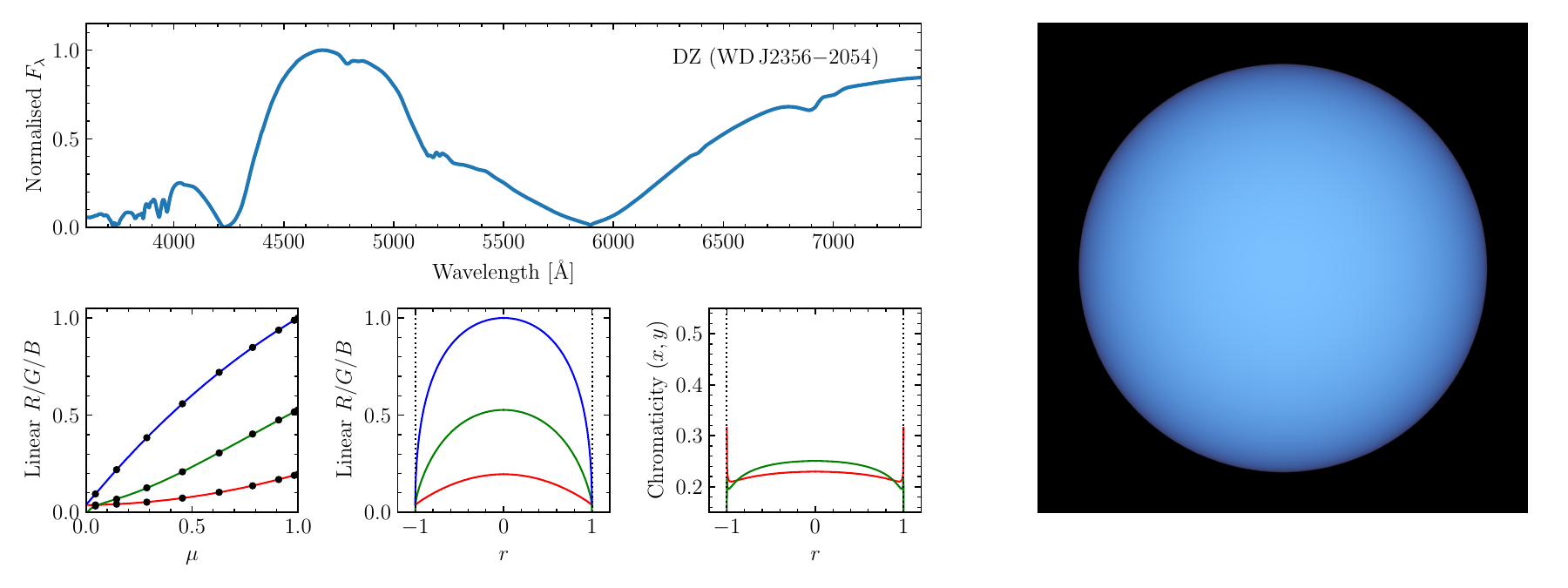}
  \caption{\label{fig:visDZ2}
    Visualisations for three DZ white dwarf models. Panels to the left of the main images have the same meaning
    as in Figure\,\ref{fig:visDA}.
    }
\end{figure*}

\onecolumn

\section{Additional tables}
\FloatBarrier
\label{sec:app_tab}

\begin{table*}
  \begin{centering}
    \caption{\label{tab:DA}
    Colour information for pure hydrogen (DA) models. $(x, y)$ is the chromaticity
    and CCT is the correlated colour temperature.
    }
    \begin{tabular}{rcrccrcrc}
      \hline
      \Teff\ [K] & ($x,y$) & CCT [K] & Hexcode & & \Teff\ [K] & ($x,y$) & CCT [K] & Hexcode \\
      \hline
140\,000 & (0.242, 0.238) & 102\,163 & \#98B3FF &    &  13\,500 & (0.266, 0.277) & 13\,046 & \#B6D1FF \\
130\,000 & (0.243, 0.238) & 97\,064 & \#98B4FF &    &  13\,000 & (0.267, 0.279) & 12\,605 & \#B8D2FF \\
120\,000 & (0.243, 0.238) & 91\,918 & \#98B4FF &    &  12\,500 & (0.268, 0.281) & 12\,226 & \#BAD4FF \\
110\,000 & (0.243, 0.239) & 86\,694 & \#98B4FF &    &  12\,000 & (0.270, 0.282) & 11\,837 & \#BCD5FF \\
100\,000 & (0.243, 0.239) & 81\,394 & \#98B4FF &    &  11\,500 & (0.272, 0.284) & 11\,428 & \#BFD6FF \\
 90\,000 & (0.243, 0.240) & 76\,020 & \#99B5FF &    &  11\,000 & (0.274, 0.286) & 11\,029 & \#C2D8FF \\
 85\,000 & (0.243, 0.240) & 73\,230 & \#99B5FF &    &  10\,500 & (0.276, 0.287) & 10\,633 & \#C5D9FF \\
 80\,000 & (0.244, 0.240) & 70\,387 & \#99B5FF &    &  10\,000 & (0.278, 0.289) & 10\,236 & \#C9DAFF \\
 75\,000 & (0.244, 0.240) & 67\,487 & \#99B5FF &    &     9500 & (0.280, 0.291) & 9844 & \#CCDCFF \\
 70\,000 & (0.244, 0.241) & 64\,513 & \#9AB5FF &    &     9000 & (0.283, 0.294) & 9441 & \#D0DEFF \\
 65\,000 & (0.244, 0.241) & 61\,440 & \#9AB6FF &    &     8500 & (0.286, 0.296) & 9018 & \#D5E0FF \\
 60\,000 & (0.244, 0.241) & 58\,248 & \#9AB6FF &    &     8000 & (0.290, 0.300) & 8569 & \#DBE3FF \\
 55\,000 & (0.245, 0.242) & 54\,895 & \#9AB6FF &    &     7500 & (0.294, 0.304) & 8087 & \#E1E7FF \\
 50\,000 & (0.245, 0.243) & 51\,324 & \#9BB7FF &    &     7000 & (0.300, 0.310) & 7562 & \#EAECFF \\
 45\,000 & (0.245, 0.243) & 47\,433 & \#9BB7FF &    &     6500 & (0.307, 0.317) & 6978 & \#F5F2FF \\
 40\,000 & (0.246, 0.244) & 43\,041 & \#9CB8FF &    &     6000 & (0.316, 0.326) & 6326 & \#FFF7FA \\
 35\,000 & (0.247, 0.246) & 37\,806 & \#9DB9FF &    &     5500 & (0.331, 0.342) & 5571 & \#FFF0E3 \\
 30\,000 & (0.249, 0.250) & 30\,368 & \#9FBCFF &    &     5250 & (0.342, 0.353) & 5143 & \#FFEBD4 \\
 25\,000 & (0.252, 0.255) & 23\,921 & \#A3C0FF &    &     5000 & (0.354, 0.364) & 4746 & \#FFE6C3 \\
 20\,000 & (0.256, 0.261) & 19\,009 & \#A8C5FF &    &     4750 & (0.366, 0.374) & 4406 & \#FFE0B4 \\
 17\,000 & (0.259, 0.267) & 16\,356 & \#ADC9FF &    &     4500 & (0.378, 0.384) & 4110 & \#FFDAA5 \\
 16\,500 & (0.260, 0.268) & 15\,904 & \#AECAFF &    &     4250 & (0.390, 0.392) & 3855 & \#FFD497 \\
 16\,000 & (0.260, 0.269) & 15\,440 & \#AFCBFF &    &     4000 & (0.402, 0.399) & 3643 & \#FFCF8B \\
 15\,500 & (0.261, 0.271) & 14\,971 & \#B0CCFF &    &     3750 & (0.412, 0.404) & 3470 & \#FFCA80 \\
 15\,000 & (0.262, 0.272) & 14\,494 & \#B1CDFF &    &     3500 & (0.421, 0.409) & 3326 & \#FFC677 \\
 14\,500 & (0.263, 0.274) & 14\,010 & \#B3CEFF &    &     3250 & (0.429, 0.413) & 3209 & \#FFC26F \\
 14\,000 & (0.264, 0.275) & 13\,524 & \#B4D0FF &    &     3000 & (0.435, 0.416) & 3120 & \#FFBF68 \\
      \hline
    \end{tabular}
  \end{centering}
\end{table*}

\begin{table*}
  \begin{centering}
    \caption{\label{tab:DB}
    Colour information for pure helium (DB) models.
    }
    \begin{tabular}{rcrccrcrc}
      \hline
      \Teff\ [K] & ($x,y$) & CCT [K] & Hexcode & & \Teff\ [K] & ($x,y$) & CCT [K] & Hexcode \\
      \hline
 40\,000 & (0.247, 0.248) & 33\,532 & \#9DBBFF &    &  15\,000 & (0.261, 0.266) & 15\,803 & \#B1C8FF \\
 39\,000 & (0.248, 0.249) & 32\,372 & \#9EBCFF &    &  14\,000 & (0.263, 0.268) & 15\,004 & \#B3C9FF \\
 38\,000 & (0.248, 0.249) & 31\,287 & \#9EBCFF &    &  13\,000 & (0.265, 0.271) & 14\,050 & \#B7CBFF \\
 37\,000 & (0.248, 0.250) & 30\,258 & \#9FBCFF &    &  12\,000 & (0.268, 0.274) & 12\,996 & \#BBCEFF \\
 36\,000 & (0.249, 0.251) & 29\,276 & \#9FBDFF &    &  11\,000 & (0.272, 0.278) & 11\,882 & \#C0D1FF \\
 35\,000 & (0.249, 0.251) & 28\,332 & \#A0BDFF &    &  10\,000 & (0.277, 0.284) & 10\,755 & \#C7D6FF \\
 34\,000 & (0.249, 0.252) & 27\,419 & \#A0BEFF &    &     9500 & (0.280, 0.287) & 10\,200 & \#CBD8FF \\
 33\,000 & (0.250, 0.253) & 26\,520 & \#A1BEFF &    &     9000 & (0.283, 0.291) & 9662 & \#D0DBFF \\
 32\,000 & (0.250, 0.253) & 25\,645 & \#A1BFFF &    &     8500 & (0.286, 0.294) & 9144 & \#D5DEFF \\
 31\,000 & (0.251, 0.254) & 24\,782 & \#A2C0FF &    &     8000 & (0.290, 0.299) & 8638 & \#DAE2FF \\
 30\,000 & (0.251, 0.255) & 23\,930 & \#A2C0FF &    &     7500 & (0.294, 0.303) & 8125 & \#E1E6FF \\
 29\,000 & (0.252, 0.256) & 23\,089 & \#A3C1FF &    &     7000 & (0.299, 0.309) & 7591 & \#E9EBFF \\
 28\,000 & (0.252, 0.257) & 22\,272 & \#A4C2FF &    &     6500 & (0.306, 0.316) & 7032 & \#F4F2FF \\
 27\,000 & (0.253, 0.258) & 21\,505 & \#A4C2FF &    &     6000 & (0.314, 0.325) & 6438 & \#FFF8FD \\
 26\,000 & (0.253, 0.259) & 20\,834 & \#A5C3FF &    &     5500 & (0.326, 0.336) & 5791 & \#FFF1EA \\
 25\,000 & (0.254, 0.260) & 20\,310 & \#A6C4FF &    &     5250 & (0.334, 0.343) & 5438 & \#FFEDDF \\
 24\,000 & (0.254, 0.260) & 19\,939 & \#A6C4FF &    &     5000 & (0.344, 0.352) & 5061 & \#FFE8D1 \\
 23\,000 & (0.254, 0.261) & 19\,648 & \#A7C4FF &    &     4750 & (0.356, 0.362) & 4667 & \#FFE2C2 \\
 22\,000 & (0.255, 0.261) & 19\,299 & \#A7C5FF &    &     4500 & (0.370, 0.373) & 4266 & \#FFDBB0 \\
 21\,000 & (0.255, 0.262) & 18\,794 & \#A8C5FF &    &     4250 & (0.387, 0.384) & 3875 & \#FFD29D \\
 20\,000 & (0.256, 0.263) & 18\,244 & \#A9C6FF &    &     4000 & (0.406, 0.395) & 3512 & \#FFC989 \\
 19\,000 & (0.257, 0.264) & 17\,732 & \#AAC6FF &    &     3750 & (0.426, 0.403) & 3186 & \#FFBF76 \\
 18\,000 & (0.258, 0.264) & 17\,280 & \#ACC7FF &    &     3500 & (0.448, 0.410) & 2874 & \#FFB462 \\
 17\,000 & (0.259, 0.265) & 16\,867 & \#ADC7FF &    &     3250 & (0.472, 0.415) & 2569 & \#FFA84C \\
 16\,000 & (0.260, 0.265) & 16\,409 & \#AFC7FF &    &     3000 & (0.492, 0.417) & 2353 & \#FF9E3B \\
      \hline
      \newline
    \end{tabular}
  \end{centering}
\end{table*}

\begin{table*}
\setlength{\tabcolsep}{4pt}
  \begin{centering}
    \caption{\label{tab:UC}
    Colour information for IR-faint white dwarf spectra.
    }
    \begin{tabular}{ccrccccrc}
      \hline
      WD\,J & ($x,y$) & CCT [K] & Hexcode & & WD\,J & ($x,y$) & CCT [K] & Hexcode \\
      \hline
003908.38$+$303539.76 & (0.281, 0.320) &    8626 & \#CBF6FF &    & 144440.02$+$631924.43 & (0.295, 0.318) &    7753 & \#E1F4FF \\
085443.36$+$350353.26 & (0.387, 0.400) &    3985 & \#FFDD96 &    & 170320.13$+$271106.89 & (0.336, 0.355) &    5380 & \#FFF5DB \\
094722.99$+$445948.59 & (0.338, 0.347) &    5281 & \#FFEBD9 &    & 172748.75$+$080819.54 & (0.353, 0.362) &    4743 & \#FFE4C4 \\
100103.50$+$390341.05 & (0.321, 0.348) &    6001 & \#FAFFEC &    & 223954.13$+$001847.11 & (0.331, 0.339) &    5570 & \#FFEDE3 \\
125106.16$+$440303.04 & (0.262, 0.286) & 12\,738 & \#AED9FF &    & 225817.72$+$012811.62 & (0.304, 0.330) &    6969 & \#F0FFFE \\
133739.41$+$000142.66 & (0.321, 0.342) &    6003 & \#FFFEF0 &    & 231130.15$+$283444.82 & (0.300, 0.334) &    7154 & \#E6FFFA \\
140237.32$+$080519.16 & (0.287, 0.317) &    8282 & \#D5F4FF &    & 234931.17$-$080245.64 & (0.302, 0.322) &    7211 & \#EDF8FF \\
140324.75$+$453333.02 & (0.278, 0.318) &    8920 & \#C5F5FF &    &  & & &  \\
      \hline
    \end{tabular}
  \end{centering}
\end{table*}

\begin{table*}
\setlength{\tabcolsep}{4pt}
  \begin{centering}
    \caption{\label{tab:DQ}
    Colour information for DQ spectra.
    }
    \begin{tabular}{ccrccccrc}
      \hline
      WD\,J & ($x,y$) & CCT [K] & Hexcode & & WD\,J & ($x,y$) & CCT [K] & Hexcode \\
      \hline
013655.67$+$200429.88 & (0.295, 0.281) &    8781 & \#E3D1FF &    & 114136.50$+$383611.86 & (0.358, 0.303) &    4109 & \#FFB2CC \\
014519.42$+$231753.94 & (0.286, 0.288) &    9450 & \#D5D8FF &    & 115933.13$+$130031.54 & (0.316, 0.289) &    6654 & \#FFD4FD \\
032054.12$-$071625.45 & (0.320, 0.293) &    6319 & \#FFD2F7 &    & 120154.71$+$340055.34 & (0.325, 0.318) &    5900 & \#FFE3EE \\
074008.46$+$181047.41 & (0.296, 0.300) &    8067 & \#E4E2FF &    & 123313.51$+$082402.65 & (0.310, 0.312) &    6839 & \#F9EEFF \\
080558.85$+$072449.19 & (0.341, 0.338) &    5111 & \#FFE0D7 &    & 123752.23$+$415624.69 & (0.338, 0.294) &    5135 & \#FFBFE2 \\
080843.15$+$464028.64 & (0.377, 0.327) &    3649 & \#FFB4B6 &    & 124739.13$+$064604.30 & (0.407, 0.376) &    3335 & \#FFBC90 \\
082005.03$+$520453.60 & (0.317, 0.295) &    6532 & \#FFD8FB &    & 130945.66$+$444541.02 & (0.296, 0.292) &    8275 & \#E5DBFF \\
083231.55$-$040806.97 & (0.323, 0.284) &    6200 & \#FFC7F4 &    & 131147.07$+$292349.67 & (0.366, 0.303) &    3762 & \#FFACC4 \\
085239.66$+$042804.48 & (0.274, 0.270) & 12\,517 & \#C4CAFF &    & 133359.84$+$001655.03 & (0.387, 0.306) &    3063 & \#FF9DB3 \\
085626.95$+$451337.01 & (0.275, 0.262) & 13\,386 & \#C6C3FF &    & 135516.62$+$363612.70 & (0.285, 0.284) &    9775 & \#D4D5FF \\
085914.71$+$325712.16 & (0.278, 0.275) & 11\,251 & \#CBCDFF &    & 135628.24$-$000941.34 & (0.311, 0.302) &    6911 & \#FAE3FF \\
085935.89$+$601616.14 & (0.287, 0.296) &    9002 & \#D6E0FF &    & 145725.28$+$210747.81 & (0.322, 0.319) &    6026 & \#FFE7F1 \\
090208.37$+$201051.57 & (0.384, 0.320) &    3346 & \#FFA9B2 &    & 150126.78$+$210056.68 & (0.329, 0.314) &    5674 & \#FFDAE9 \\
093537.00$+$002422.00 & (0.371, 0.316) &    3742 & \#FFB1BE &    & 155413.52$+$033634.74 & (0.304, 0.313) &    7224 & \#F0EFFF \\
095051.75$+$122758.93 & (0.427, 0.389) &    3054 & \#FFB67B &    & 161414.19$+$172901.23 & (0.355, 0.334) &    4541 & \#FFCEC9 \\
101141.58$+$284559.07 & (0.463, 0.347) &    2147 & \#FF8775 &    & 180302.55$+$232044.14 & (0.416, 0.374) &    3127 & \#FFB58A \\
102801.78$+$351258.64 & (0.360, 0.333) &    4332 & \#FFC8C4 &    & 183500.11$+$642917.66 & (0.316, 0.286) &    6688 & \#FFD1FD \\
111341.34$+$014642.06 & (0.360, 0.315) &    4160 & \#FFBAC8 &    & 231030.26$-$005745.76 & (0.294, 0.289) &    8610 & \#E1D9FF \\
      \hline
    \end{tabular}
  \end{centering}
\end{table*}

\begin{table*}
\setlength{\tabcolsep}{4pt}
  \begin{centering}
    \caption{\label{tab:DZ}
    Colour information for DZ models.
    }
    \begin{tabular}{ccrccccrc}
    
      \hline
      WD\,J & ($x,y$) & CCT [K] & Hexcode & & WD\,J & ($x,y$) & CCT [K] & Hexcode \\
      \hline
004757.08$+$162836.45 & (0.308, 0.321) &    6830 & \#F7F7FF &    & 115748.35$+$613845.96 & (0.295, 0.305) &    7991 & \#E3E7FF \\
015848.99$-$094225.19 & (0.309, 0.326) &    6760 & \#F8FCFF &    & 115809.40$+$184557.49 & (0.291, 0.292) &    8705 & \#DDDCFF \\
025206.01$-$040130.28 & (0.290, 0.300) &    8531 & \#DBE3FF &    & 132144.04$-$023751.47 & (0.322, 0.349) &    5949 & \#FBFFEA \\
073635.21$+$411828.68 & (0.369, 0.394) &    4430 & \#FFEBAB &    & 133001.17$+$643523.69 & (0.332, 0.404) &    5531 & \#DDFFB8 \\
074414.64$+$464912.63 & (0.361, 0.421) &    4744 & \#F6FFA3 &    & 133059.42$+$302953.65 & (0.313, 0.328) &    6492 & \#FFFEFF \\
080626.70$+$305555.81 & (0.300, 0.322) &    7291 & \#EBF8FF &    & 135123.88$+$264546.53 & (0.331, 0.374) &    5552 & \#F3FFD2 \\
080740.69$+$493059.73 & (0.365, 0.391) &    4523 & \#FFEDB0 &    & 143007.16$-$015129.66 & (0.320, 0.338) &    6086 & \#FFFDF3 \\
085100.24$+$154302.04 & (0.308, 0.336) &    6712 & \#F1FFF8 &    & 150028.03$+$231554.04 & (0.300, 0.310) &    7508 & \#EBECFF \\
085217.60$+$340211.27 & (0.325, 0.352) &    5804 & \#FDFFE7 &    & 153505.81$+$124745.20 & (0.317, 0.332) &    6237 & \#FFFAF7 \\
090146.88$+$075206.86 & (0.302, 0.316) &    7298 & \#EDF2FF &    & 154022.80$+$535239.89 & (0.298, 0.309) &    7664 & \#E8EBFF \\
090244.75$+$100459.69 & (0.292, 0.309) &    8144 & \#DDEBFF &    & 154201.78$+$465020.25 & (0.315, 0.329) &    6378 & \#FFFCFB \\
091621.31$+$254029.10 & (0.353, 0.381) &    4817 & \#FFF3BF &    & 154510.32$+$523618.18 & (0.313, 0.334) &    6446 & \#FBFFFA \\
095645.15$+$591240.64 & (0.278, 0.283) & 10\,605 & \#CAD5FF &    & 163601.35$+$161907.51 & (0.387, 0.420) &    4109 & \#FFE98E \\
102438.07$+$101410.60 & (0.321, 0.335) &    6065 & \#FFF8F2 &    & 164939.23$+$223807.22 & (0.335, 0.343) &    5393 & \#FFEBDE \\
103809.19$-$003622.45 & (0.287, 0.304) &    8673 & \#D6E8FF &    & 182458.45$+$121316.82 & (0.380, 0.395) &    4131 & \#FFE09F \\
104658.11$+$132911.33 & (0.337, 0.361) &    5342 & \#FFF8D8 &    & 222503.65$+$233855.87 & (0.290, 0.301) &    8496 & \#DBE4FF \\
105533.75$+$372543.21 & (0.347, 0.393) &    5053 & \#F9FFBF &    & 223507.66$-$005607.73 & (0.303, 0.320) &    7147 & \#EFF6FF \\
113209.59$+$332353.03 & (0.309, 0.321) &    6776 & \#F9F6FF &    & 235645.10$-$205449.94 & (0.222, 0.234) & - & \#70B4FF \\
      \hline
    \end{tabular}
  \end{centering}
\end{table*}

\clearpage

\begin{longtable}{clcrccrccc}
    \caption{\label{tab:twenty}
    Colour information for white dwarfs within 20\,pc of the Sun. Surface gravities are
    in cgs units.
    } \\
    \hline
    WD\,J & Alt. name & SpT & \Teff\ [K] & $\log g$ [dex] & ($x,y$) & CCT [K] & Hexcode & ref & note \\
    \hline
    \endfirsthead
    \caption{Continued.} \\
    \hline
    WD\,J & Alt. name & SpT & \Teff\ [K] & $\log g$ [dex] & ($x,y$) & CCT [K] & Hexcode & ref & note \\
    \hline
    \endhead
    \hline
    \endfoot 
    \hline \\
    \caption*{References:
    (1) \citet{obrienetal24-1},
    (2) \citet{giammicheleetal12-1},
    (3) \citet{obrienetal26-2},
    (4) \citet{farihietal13-3},
    (5) \citet{vincentetal24-1},
    (6) \citet{bergeronetal97-1},
    (7) \citet{obrienetal26-1},
    (8) \citet{tremblayetal17-1},
    (9) \citet{coutuetal19-1},
    (10) \citet{blouinetal19-4},
    (11) \citet{hollandsetal22-1},
    (12) \citet{blouinetal18-1}.
    \newline
    Notes:
    (a) Weak carbon/metal features. Colour calculations were performed with pure
    helium model grids.
    (b) Object is an unresolved double-degenerate binary. Parameters for
    each component are averages from the provided reference.
    (c) Object is absent from \emph{Gaia} DR3, with parameters adopted from the
    accompanying reference.
    (d) White dwarf is part of post-common-envelope binary with atmospheric
    parameters determined from UV spectroscopy, and a DA spectral type assumed
    for colour calculations.
    (e) Spectrum was obtained from the SDSS.
    }
    \endlastfoot 
000210.72$-$430955.39 & GJ\,915 & DA & 8428 & 8.354 & (0.287, 0.297) & 8926 & \#D6E1FF & 1 &  \\
000240.09$-$341339.59 & LHS\,1008 & DC & 6332 & 8.143 & (0.309, 0.319) & 6826 & \#F8F4FF & 1 &  \\
000720.55$+$123021.16 & LP\,464$-$57 & DCP & 5223 & 8.274 & (0.345, 0.356) & 5027 & \#FFEACF & 1 &  \\
001214.75$+$502520.74 & LHS\,1038 & DA & 6483 & 8.237 & (0.307, 0.317) & 6933 & \#F6F3FF & 1 &  \\
001349.89$-$714954.26 & L\,50$-$73 & DAH & 6275 & 7.874 & (0.311, 0.321) & 6677 & \#FCF7FF & 1 &  \\
001412.79$-$131101.09 & G\,158$-$45 & DAH & 5871 & 8.212 & (0.321, 0.331) & 6066 & \#FFF5F2 & 1 &  \\
004126.03$-$222102.29 & LHS\,1126 & DQ & 5368 & 7.990 & (0.329, 0.344) & 5658 & \#FFF4E5 & 2 &  \\
004909.90$+$052318.99 & van Maanen 2 & DZA & 5830 & 8.09 & (0.305, 0.320) & 7067 & \#F2F5FF & 3 &  \\
011800.08$+$161020.56 & LHS\,1227 & DQ & 9278 & 8.078 & (0.281, 0.289) & 9957 & \#CDD9FF & 1 & a \\
012403.99$-$424038.43 & LP\,991$-$16 & DAH & 6080 & 7.583 & (0.314, 0.324) & 6466 & \#FFF8FD & 1 &  \\
012524.47$-$260044.45 & GJ\,1039 & DC & 6941 & 8.227 & (0.300, 0.310) & 7527 & \#EAECFF & 1 &  \\
013759.39$-$045944.67 & L\,870$-$2 & DA+DA & 7454 & 7.906 & (0.295, 0.305) & 8039 & \#E2E7FF & 3 & b \\
                      &            &       & 7162 & 7.809 & (0.298, 0.308) & 7730 & \#E7EAFF &   &   \\
014300.98$-$671830.35 & L\,88$-$59 & DAZ & 6350 & 7.981 & (0.310, 0.320) & 6768 & \#F9F5FF & 1 &  \\
015151.14$+$642552.55 & GJ\,3118 & DA & 8794 & 8.098 & (0.285, 0.295) & 9247 & \#D3DFFF & 1 &  \\
015202.96$+$470006.66 & GD\,279 & DA & 14\,293 & 7.999 & (0.264, 0.274) & 13\,810 & \#B3CFFF & 1 &  \\
021026.11$-$504926.33 & GJ\,86B & DC & 8180 & 8.020 & (0.288, 0.297) & 8812 & \#D8E1FF & 4 & c \\
021120.84$+$395521.52 & LHS\,151 & DAZ & 7196 & 7.933 & (0.298, 0.307) & 7760 & \#E6EAFF & 1 &  \\
021228.98$-$080411.00 & LP\,649$-$67 & DA & 8467 & 7.893 & (0.287, 0.297) & 9000 & \#D5E0FF & 1 &  \\
021348.83$-$334530.03 & LP\,941$-$19 & DAZ & 5618 & 8.020 & (0.327, 0.338) & 5725 & \#FFF1E8 & 5 &  \\
023237.91$-$141151.77 & LHS\,1415 & DA & 5393 & 7.986 & (0.335, 0.346) & 5381 & \#FFEEDD & 1 &  \\
023521.79$-$240047.12 & LP\,830$-$14 & DAH & 4871 & 7.529 & (0.357, 0.366) & 4645 & \#FFE4BF & 1 &  \\
024836.43$+$542322.46 & LHS\,1446 & DAZ & 4942 & 8.018 & (0.356, 0.367) & 4659 & \#FFE5C0 & 1 &  \\
031031.02$-$683603.38 & LB\,3303 & DA & 15\,866 & 8.077 & (0.261, 0.270) & 15\,107 & \#AFCCFF & 1 &  \\
032511.05$-$014915.05 & LHS\,1547 & DAZH & 5165 & 8.048 & (0.346, 0.357) & 4988 & \#FFE9CE & 1 &  \\
034434.85$+$182609.82 & GJ\,151 & DQ & 6837 & 8.097 & (0.303, 0.308) & 7374 & \#EEE9FF & 6 &  \\
040026.68$+$081406.93 & GJ\,3259 & DA & 5461 & 7.934 & (0.332, 0.343) & 5514 & \#FFEFE1 & 1 &  \\
041521.80$-$073929.20 & 40 Eridani B & DA & 16\,265 & 7.916 & (0.260, 0.268) & 15\,880 & \#AECAFF & 1 &  \\
041630.04$-$591757.19 & $\epsilon$ Reticuli B & DA & 14\,271 & 7.824 & (0.263, 0.273) & 14\,177 & \#B2CEFF & 1 &  \\
042553.73$+$121148.34 & LP\,475$-$70 & DC & 6153 & 8.213 & (0.312, 0.322) & 6621 & \#FDF7FF & 1 &  \\
043112.57$+$585841.29 & Stein 2051 B & DC & 7011 & 8.098 & (0.299, 0.309) & 7604 & \#E9EBFF & 1 &  \\
043644.90$+$270951.52 & G\,39$-$27 & DA & 5534 & 8.023 & (0.330, 0.341) & 5610 & \#FFF0E4 & 1 &  \\
043747.41$-$084910.62 & LHS\,194 & DQ & 6601 & 8.050 & (0.308, 0.310) & 6984 & \#F6ECFF & 6 &  \\
050552.46$-$172243.48 & GJ\,3329 & DAH & 5345 & 7.864 & (0.337, 0.347) & 5330 & \#FFEDDB & 1 &  \\
053344.80$+$015643.45 & GJ\,207.1 B & - & 6100 & 8.0 & (0.314, 0.325) & 6445 & \#FFF8FD & 7 & d \\
055119.48$-$001021.11 & G\,99$-$37 & DQ & 6053 & 8.122 & (0.323, 0.330) & 5967 & \#FFF1F0 & 6 &  \\
055443.04$-$103521.34 & WD\,0552$-$106 & DZ & 6578 & 8.177 & (0.305, 0.315) & 7120 & \#F2F1FF & 1 &  \\
055509.53$-$041007.07 & LP\,658$-$2 & DZ & 4642 & 8.098 & (0.362, 0.378) & 4531 & \#FFE6B6 & 3 &  \\
055625.46$+$052148.44 & G\,99$-$47 & DAH & 5729 & 8.151 & (0.325, 0.336) & 5852 & \#FFF3EC & 1 &  \\
060529.40$+$604922.46 & LHS\,1817 B & - & 6300 & 8.0 & (0.311, 0.321) & 6699 & \#FBF6FF & 7 & d \\
064509.30$-$164300.72 & Sirius B & DA & 25\,970 & 8.570 & (0.252, 0.256) & 22\,628 & \#A4C1FF & 8 &  \\
064722.15$+$023108.91 & GJ\,3410 & DA & 7109 & 8.594 & (0.299, 0.309) & 7650 & \#E8EBFF & 1 &  \\
064737.99$+$373057.06 & LAWD\,23 & DA & 21\,394 & 8.145 & (0.255, 0.260) & 19\,627 & \#A7C4FF & 1 &  \\
065705.91$-$390935.68 & L\,454$-$9 & DA & 6370 & 8.080 & (0.309, 0.319) & 6785 & \#F9F5FF & 1 &  \\
070051.66$+$315743.83 & LHS\,1889 & DA & 4873 & 7.907 & (0.359, 0.369) & 4582 & \#FFE3BC & 1 &  \\
070852.28$-$670631.43 & WD\,0708$-$670 & DCP & 4862 & 7.860 & (0.359, 0.369) & 4574 & \#FFE3BC & 1 &  \\
073046.96$+$481006.27 & G\,107$-$70 B & DA & 4926 & 8.100 & (0.358, 0.368) & 4623 & \#FFE4BE & 2 & c \\
073046.99$+$481005.67 & G\,107$-$70 A & DA & 4934 & 7.890 & (0.356, 0.366) & 4671 & \#FFE5C0 & 2 & c \\
073330.88$+$640927.44 & GJ\,1098 & DA & 5140 & 7.824 & (0.346, 0.356) & 5000 & \#FFE9CE & 1 &  \\
073917.88$+$051326.80 & Procyon B & DQZ & 7585 & 7.960 & (0.289, 0.296) & 8799 & \#D9E0FF & 2 & c \\
074020.79$-$172449.16 & LAWD\,25 & DZA & 7450 & 7.972 & (0.289, 0.297) & 8756 & \#D9E1FF & 3 &  \\
074538.43$-$335551.34 & GJ\,288 B & DC & 4293 & 7.798 & (0.388, 0.390) & 3911 & \#FFD69A & 1 &  \\
075014.58$+$071148.92 & GJ\,1102 B & DC & 4280 & 7.783 & (0.388, 0.390) & 3899 & \#FFD59A & 1 &  \\
075015.34$+$071137.10 & GJ\,1102 A & DC & 4680 & 7.894 & (0.369, 0.376) & 4332 & \#FFDFB0 & 1 &  \\
075308.14$-$674731.38 & L\,97$-$12 & DA & 5630 & 7.948 & (0.327, 0.337) & 5757 & \#FFF2E9 & 1 &  \\
075356.62$-$252401.48 & WD\,0751$-$252 & DA & 4939 & 7.897 & (0.356, 0.366) & 4677 & \#FFE5C0 & 1 &  \\
080653.75$-$661816.70 & L\,97$-$3 & DQ & 10\,393 & 8.041 & (0.275, 0.282) & 11\,165 & \#C5D4FF & 1 & a \\
081227.07$-$352943.32 & WD\,0810$-$353 & DAH & 6239 & 8.178 & (0.312, 0.322) & 6597 & \#FDF8FF & 1 &  \\
081411.17$+$484529.81 & G\,111$-$64 & DC & 6709 & 8.119 & (0.303, 0.313) & 7257 & \#EFEFFF & 1 &  \\
081840.26$-$311020.33 & WD\,0816$-$310 & DZ & 6754 & 8.263 & (0.295, 0.311) & 7829 & \#E3EEFF & 9 &  \\
082126.70$-$670320.02 & WD\,0820$-$668 & DA & 4793 & 7.902 & (0.363, 0.372) & 4475 & \#FFE1B7 & 1 &  \\
084132.43$-$325632.92 & GJ\,318 & DA & 9086 & 7.770 & (0.282, 0.293) & 9563 & \#CFDDFF & 1 &  \\
084248.46$-$134713.21 & LP\,726$-$1 & DZ & 5388 & 8.174 & (0.335, 0.365) & 5424 & \#FFFFDA & 9 &  \\
085912.92$-$005842.86 & LP\,606$-$32 & DC & 4997 & 7.947 & (0.353, 0.364) & 4753 & \#FFE6C4 & 1 &  \\
091556.08$+$532523.86 & G\,195$-$19 & DCP & 7155 & 8.217 & (0.298, 0.307) & 7757 & \#E6E9FF & 1 &  \\
100149.36$+$144123.79 & G\,42$-$33 & DC & 6778 & 8.206 & (0.302, 0.312) & 7338 & \#EEEEFF & 1 &  \\
101141.58$+$284559.07 & LHS\,2229 & DQpec & 4335 & 8.211 & (0.463, 0.347) & 2151 & \#FF8775 & 10 & e \\
101201.87$-$184333.27 & WD\,1009$-$184 & DZ & 6394 & 8.183 & (0.305, 0.319) & 7048 & \#F2F4FF & 9 &  \\
102309.06$+$632741.49 & LP\,62$-$147 & DA & 6801 & 8.033 & (0.303, 0.312) & 7313 & \#EEEEFF & 1 &  \\
103702.48$+$711058.70 & LHS\,285 & DC & 4693 & 8.155 & (0.369, 0.377) & 4314 & \#FFDEAF & 1 &  \\
103855.57$-$204056.83 & LHS\,2293 & DQ & 5754 & 8.557 & (0.358, 0.302) & 4092 & \#FFB1CC & 2 &  \\
104539.04$-$190645.03 & GJ\,401 B & DQ & 5736 & 7.875 & (0.325, 0.324) & 5869 & \#FFE8ED & 6 &  \\
105735.13$-$073123.18 & LHS\,2333 & DC & 7096 & 8.245 & (0.298, 0.308) & 7696 & \#E7EAFF & 1 &  \\
111827.20$-$472156.96 & WD\,1116$-$470 & DC & 5910 & 8.077 & (0.316, 0.327) & 6316 & \#FFF7F9 & 1 &  \\
112412.97$+$212135.57 & Ross\,627 & DA & 7373 & 8.037 & (0.296, 0.305) & 7944 & \#E3E8FF & 1 &  \\
113430.48$-$325002.40 & GJ\,432 B & DZ & 5178 & 8.024 & (0.333, 0.342) & 5481 & \#FFEEE0 & 3 &  \\
113705.10$+$294758.29 & GD\,140 & DA & 21\,336 & 8.517 & (0.256, 0.263) & 18\,415 & \#A9C6FF & 1 &  \\
114542.92$-$645029.46 & LAWD 37 & DQ & 8062 & 8.058 & (0.284, 0.289) & 9629 & \#D1DAFF & 6 &  \\
114734.45$-$745759.24 & WD\,1145$-$747 & DC & 3821 & 7.742 & (0.409, 0.403) & 3524 & \#FFCC84 & 1 &  \\
115052.33$+$683115.97 & WD\,1148$+$687 & DA & 6696 & 8.258 & (0.304, 0.314) & 7177 & \#F1F0FF & 1 &  \\
120526.67$-$233312.14 & LP\,852$-$7 & DAZ & 8540 & 7.959 & (0.286, 0.296) & 9055 & \#D5E0FF & 1 &  \\
121129.27$+$572417.24 & G\,197$-$47 & DAZ & 5829 & 7.982 & (0.321, 0.332) & 6044 & \#FFF4F2 & 1 &  \\
122642.02$-$661218.47 & GJ\,2092 & DA & 7582 & 7.897 & (0.294, 0.303) & 8166 & \#E0E6FF & 1 &  \\
123849.78$-$494800.22 & GJ\,2095 & DA & 11\,359 & 8.688 & (0.274, 0.287) & 10\,813 & \#C3D9FF & 1 &  \\
130009.06$+$032841.06 & GJ\,492 & DA & 5650 & 8.190 & (0.327, 0.339) & 5724 & \#FFF2E8 & 1 &  \\
130841.20$+$850228.16 & G\,256$-$7 & DAH & 5254 & 8.077 & (0.342, 0.353) & 5128 & \#FFEBD3 & 1 &  \\
131256.47$-$472807.91 & GJ\,3770 & DC & 3855 & 7.695 & (0.407, 0.402) & 3549 & \#FFCC85 & 1 &  \\
131924.74$-$214754.87 & LP\,854$-$50 & DA & 5882 & 8.618 & (0.322, 0.333) & 5998 & \#FFF5F0 & 1 &  \\
131925.56$-$782328.17 & LAWD\,45 & DAH & 5542 & 8.129 & (0.330, 0.342) & 5590 & \#FFF0E4 & 1 &  \\
133013.64$-$083429.47 & GJ\,515 & DA & 14\,563 & 7.971 & (0.263, 0.273) & 14\,139 & \#B2CEFF & 1 &  \\
133631.85$+$034045.94 & Wolf 489 & DA & 4858 & 7.821 & (0.359, 0.369) & 4575 & \#FFE3BC & 1 &  \\
134121.80$+$050045.85 & WD\,1338$+$052 & DC & 4022 & 7.746 & (0.400, 0.398) & 3671 & \#FFD08D & 1 &  \\
134803.01$+$233446.44 & LP\,380$-$5 & DC & 4576 & 7.767 & (0.373, 0.380) & 4222 & \#FFDCAB & 1 &  \\
135315.53$-$091633.37 & GJ\,3814 & DAH & 8881 & 8.325 & (0.284, 0.295) & 9295 & \#D2DFFF & 1 &  \\
141159.17$-$592044.99 & WD\,1408$-$591 & DA & 6653 & 8.108 & (0.305, 0.315) & 7137 & \#F1F0FF & 1 &  \\
144725.34$-$174215.75 & LHS\,378 & DC & 4980 & 8.448 & (0.357, 0.368) & 4639 & \#FFE4BE & 1 &  \\
153505.81$+$124745.20 & G\,137$-$24 & DZ & 5790 & 7.985 & (0.317, 0.332) & 6237 & \#FFFAF7 & 11 &  \\
154730.02$-$375508.46 & LAWD\,60 & DA & 10\,392 & 8.026 & (0.276, 0.288) & 10\,525 & \#C6D9FF & 1 &  \\
162333.84$-$391346.16 & WD\,1620$-$391 & DA & 24\,073 & 8.006 & (0.252, 0.256) & 22\,759 & \#A4C1FF & 1 &  \\
162825.00$+$364615.85 & Ross\,640 & DZ & 8964 & 8.224 & (0.281, 0.289) & 9906 & \#CEDAFF & 12 &  \\
163233.17$+$085122.67 & G\,138$-$38 & DA & 5594 & 8.016 & (0.328, 0.339) & 5693 & \#FFF1E7 & 1 &  \\
163421.55$+$571008.87 & LHS\,422 & DQ & 6126 & 8.020 & (0.313, 0.324) & 6516 & \#FFF9FF & 6 &  \\
163501.43$+$431736.35 & G\,180$-$63 & DAZ & 6516 & 8.153 & (0.307, 0.317) & 6982 & \#F5F2FF & 1 &  \\
164825.63$+$590322.66 & G\,226$-$29 & DA & 12\,182 & 8.285 & (0.270, 0.283) & 11\,669 & \#BDD6FF & 1 &  \\
170641.36$-$264334.71 & WD\,1703$-$267 & DAH & 6128 & 8.343 & (0.315, 0.325) & 6433 & \#FFF8FC & 1 &  \\
170807.97$+$025737.08 & GJ\,1211 & DZ & 6720 & 8.242 & (0.298, 0.310) & 7691 & \#E7ECFF & 9 &  \\
170931.54$+$434052.77 & G\,203$-$47 B & - & 5286 & 8.0 & (0.340, 0.351) & 5199 & \#FFECD6 & 7 & d \\
174736.82$-$543631.16 & WD\,1743$-$545 & DC & 4358 & 7.821 & (0.384, 0.388) & 3975 & \#FFD79E & 1 &  \\
174807.99$+$705235.92 & G\,240$-$72 & DXP & 5137 & 8.060 & (0.338, 0.347) & 5274 & \#FFEBD9 & 1 &  \\
174950.15$+$824626.06 & LHS\,56 & DA & 7188 & 7.868 & (0.298, 0.307) & 7754 & \#E6EAFF & 1 &  \\
181706.50$+$132824.99 & WD\,1814$+$134 & DA & 4873 & 7.803 & (0.358, 0.368) & 4600 & \#FFE3BD & 1 &  \\
182119.83$+$610107.38 & G\,227$-$28 & DA & 4823 & 7.765 & (0.361, 0.370) & 4537 & \#FFE2BA & 1 &  \\
182404.55$-$130842.29 & GJ\,2135 & DAZ & 6079 & 8.022 & (0.315, 0.325) & 6417 & \#FFF8FC & 1 &  \\
182524.24$+$113557.34 & WD\,1823$+$116 & DA & 4883 & 7.885 & (0.359, 0.368) & 4599 & \#FFE3BD & 1 &  \\
183020.27$+$544727.21 & G\,227$-$35 & DXP & 6711 & 8.473 & (0.304, 0.314) & 7182 & \#F1F0FF & 1 &  \\
190010.25$+$703951.42 & GJ\,742 & DAH & 12\,540 & 8.671 & (0.271, 0.284) & 11\,547 & \#BDD6FF & 1 &  \\
191858.63$+$384321.48 & GJ\,1234 & DZ & 5743 & 7.939 & (0.312, 0.322) & 6589 & \#FEF8FF & 3 &  \\
192034.92$-$074000.07 & LAWD 74 & DBQA & 10\,805 & 8.051 & (0.273, 0.279) & 11\,639 & \#C2D2FF & 1 &  \\
192126.76$+$061322.71 & WD1919$+$061 & DA & 5908 & 8.147 & (0.319, 0.330) & 6142 & \#FFF5F5 & 1 &  \\
192140.42$+$144041.40 & GD\,219 & DA & 15\,162 & 8.151 & (0.263, 0.273) & 14\,254 & \#B2CEFF & 1 &  \\
193713.75$+$274318.74 & GJ\,1241 & DA & 12\,022 & 8.046 & (0.270, 0.282) & 11\,803 & \#BDD5FF & 1 &  \\
195629.23$-$010232.67 & LHS\,3501 & DAH & 7790 & 8.168 & (0.292, 0.302) & 8353 & \#DEE5FF & 1 &  \\
200502.20$+$542603.23 & Wolf\,1130 B & - & 6100 & 8.8 & (0.314, 0.325) & 6445 & \#FFF8FD & 7 & d \\
200534.87$-$105654.43 & LHS\,483 & DC & 4577 & 8.142 & (0.375, 0.381) & 4181 & \#FFDCA9 & 1 &  \\
201056.85$-$301306.63 & GJ\,2147 & DA & 15\,124 & 7.940 & (0.262, 0.271) & 14\,741 & \#B1CCFF & 1 &  \\
201231.78$-$595651.67 & WD\,2008$-$600 & DC & 4905 & 7.770 & (0.351, 0.358) & 4822 & \#FFE5C8 & 2 &  \\
202025.46$-$302714.65 & WD\,2017$-$306 & DC & 9934 & 8.045 & (0.277, 0.284) & 10\,677 & \#C8D6FF & 1 &  \\
203421.89$+$250349.75 & Wolf\,1346 & DAZ & 20\,145 & 7.932 & (0.255, 0.261) & 19\,311 & \#A8C4FF & 1 &  \\
204421.46$-$680521.36 & GJ\,2149 & DA & 16\,420 & 8.463 & (0.262, 0.272) & 14\,610 & \#B0CDFF & 1 &  \\
204906.70$+$372814.05 & LP\,284$-$13 & DAP & 14\,710 & 8.297 & (0.264, 0.276) & 13\,520 & \#B4D0FF & 1 &  \\
205020.65$+$263040.76 & G187$-$8 & DA & 4909 & 7.270 & (0.353, 0.362) & 4756 & \#FFE5C4 & 1 &  \\
205213.41$-$250415.13 & WD\,2049$-$253 & DCP & 4908 & 7.853 & (0.357, 0.367) & 4640 & \#FFE4BF & 1 &  \\
205647.78$-$045039.54 & Ross 193 B & DC & 4416 & 7.879 & (0.382, 0.386) & 4031 & \#FFD8A1 & 1 &  \\
210105.20$-$490624.27 & WD\,2058$-$493 & DA & 5118 & 7.922 & (0.347, 0.358) & 4942 & \#FFE9CC & 1 &  \\
211316.85$-$814912.88 & LAWD\,83 & DAZH & 9914 & 8.242 & (0.279, 0.290) & 10\,081 & \#CADBFF & 1 &  \\
211856.26$+$541241.24 & G\,231$-$40 & DA & 14\,785 & 7.899 & (0.262, 0.272) & 14\,511 & \#B1CDFF & 1 &  \\
214157.57$-$330029.80 & L570$-$26 & DZ & 7113 & 8.132 & (0.291, 0.306) & 8318 & \#DBE9FF & 9 &  \\
214241.01$+$205958.12 & LHS\,3703 & DQ & 8495 & 8.090 & (0.272, 0.284) & 11\,343 & \#C0D6FF & 2 &  \\
214324.09$-$065947.99 & WD\,2140$-$072 & DA & 8905 & 8.424 & (0.284, 0.295) & 9303 & \#D2DFFF & 1 &  \\
215140.11$+$591734.85 & WD\,2150$+$591 & DAH & 5105 & 7.987 & (0.348, 0.359) & 4906 & \#FFE8CA & 1 &  \\
215738.36$-$510031.52 & GJ\,841 B & DQ & 6549 & 7.969 & (0.321, 0.307) & 6178 & \#FFDEF5 & 6 &  \\
220420.86$-$751326.06 & LHS\,3752 & DA & 8875 & 8.628 & (0.284, 0.295) & 9264 & \#D2DFFF & 1 &  \\
221434.75$-$385907.27 & WD\,2211$-$392 & DA & 6063 & 8.353 & (0.316, 0.326) & 6350 & \#FFF8FA & 1 &  \\
223033.53$-$751524.30 & WD\,2226$-$755 & DC & 3911 & 7.727 & (0.405, 0.400) & 3588 & \#FFCD88 & 1 &  \\
223039.98$-$751355.24 & WD\,2226$-$754 & DC & 4194 & 7.729 & (0.392, 0.393) & 3822 & \#FFD395 & 1 &  \\
224905.55$+$223631.99 & GJ\,1273 & DA & 10\,534 & 8.662 & (0.277, 0.289) & 10\,332 & \#C7DAFF & 1 &  \\
225123.02$+$293944.49 & WD\,2150$+$591 & DA & 5635 & 7.680 & (0.326, 0.336) & 5817 & \#FFF2EB & 1 &  \\
225353.39$-$064654.49 & LP\,701$-$29 & DZ & 4041 & 7.99 & (0.408, 0.423) & 3685 & \#FFD87B & 3 &  \\
230958.53$+$550649.43 & G\,233$-$42 B & DA & 5531 & 8.107 & (0.331, 0.342) & 5581 & \#FFF0E3 & 1 &  \\
232847.64$+$051454.24 & G\,29$-$38 & DAZ & 11\,529 & 8.042 & (0.272, 0.284) & 11\,417 & \#BFD6FF & 1 &  \\
233850.74$-$074119.97 & GD\,1212 & DA & 10\,873 & 8.009 & (0.274, 0.286) & 10\,920 & \#C3D8FF & 1 &  \\
234350.72$+$323246.73 & G\,130$-$5 & DA & 12\,587 & 8.004 & (0.268, 0.280) & 12\,293 & \#BAD3FF & 1 &  \\

\end{longtable}


\bsp	
\label{lastpage}
\end{document}